\documentclass[12pt]{article}

\usepackage{amssymb,amsmath}

\newcommand{\newsection}{\setcounter{equation}{0}\section}

\newcommand{\mbf}[1]{{\boldsymbol {#1} }}

\def\appendix#1{\addtocounter{section}{1}\setcounter{equation}{0}
\renewcommand{\thesection}{\Alph{section}}
\section*{Appendix\thesection\protect\indent \parbox[t]{11.715cm} {#1}}
\addcontentsline{toc}{section}{Appendix \thesection\ \ \ #1} }
\newcommand{\zed}{{\bb Z}} 
\newcommand{\real}{{\bb R}} 
\newcommand{\zeds}{{\bbs Z}} 

\def\vp{{\mbf p}}
\def\vq{{\mbf q}}
\def\vn{{\mbf n}}
\def\vm{{\mbf m}}
\def\vd{{\mbf d}}
\def\va{{\mbf a}}
\def\vnu{{\mbf \nu}}

\font\mybb=msbm10 at 12pt
\def\bb#1{\hbox{\mybb#1}}
\font\mybbs=msbm10 at 9pt
\def\bbs#1{\hbox{\mybbs#1}}

\def\nn{\nonumber}

\newcommand{\Tr}[1]{\:{\rm Tr}\,#1}
\def\e{{\,\rm e}\,}

\newcommand{\non}{\nonumber \\}

\def\be{\begin{equation}}
\def\ee{\end{equation}}
\def\bea{\begin{eqnarray}}
\def\eea{\end{eqnarray}}
\def\bd{\begin{displaymath}}
\def\ed{\end{displaymath}}

\def\DD{{\rm D}}
\def\dd{{\rm d}}

\def\ii{{\,{\rm i}\,}}

\def\K{{{\rm K}_0}}
\def\K1{{{\rm K}_1}}

\makeatletter
\newdimen\normalarrayskip              
\newdimen\minarrayskip                 
\normalarrayskip\baselineskip
\minarrayskip\jot
\newif\ifold             \oldtrue            
\def\arraymode{\ifold\relax\else\displaystyle\fi} 
\def\@arrayskip{\ifold\baselineskip\z@\lineskip\z@
     \else
     \baselineskip\minarrayskip\lineskip2\minarrayskip\fi}
\def\@arrayclassz{\ifcase \@lastchclass \@acolampacol \or
\@ampacol \or \or \or \@addamp \or
   \@acolampacol \or \@firstampfalse \@acol \fi
\edef\@preamble{\@preamble
  \ifcase \@chnum
     \hfil$\relax\arraymode\@sharp$\hfil
     \or $\relax\arraymode\@sharp$\hfil
     \or \hfil$\relax\arraymode\@sharp$\fi}}
\def\@array[#1]#2{\setbox\@arstrutbox=\hbox{\vrule
     height\arraystretch \ht\strutbox
     depth\arraystretch \dp\strutbox
     width\z@}\@mkpream{#2}\edef\@preamble{\halign \noexpand\@halignto
\bgroup \tabskip\z@ \@arstrut \@preamble \tabskip\z@ \cr}%
\let\@startpbox\@@startpbox \let\@endpbox\@@endpbox
  \if #1t\vtop \else \if#1b\vbox \else \vcenter \fi\fi
  \bgroup \let\par\relax
  \let\@sharp##\let\protect\relax
  \@arrayskip\@preamble}
\makeatother

\newcommand{\beq}{\begin{eqnarray}}
\newcommand{\eeq}{\end{eqnarray}}

\begin{document}
\begin{titlepage}
\begin{flushright}

\baselineskip=12pt
MCTP--03--08\\ HWM--03--3\\ EMPG--03--04\\ hep-th/0302162\\
\hfill{ }\\ February 2003
\end{flushright}

\begin{center}

\baselineskip=24pt

{\Large\bf Open Wilson Lines and Group Theory of\\
Noncommutative Yang-Mills Theory\\ in Two Dimensions}

\baselineskip=14pt

\vspace{1cm}

{\bf L.D. Paniak}
\\[3mm]
{\it Michigan Center for Theoretical Physics\\ University of
Michigan\\ Ann Arbor, Michigan 48109-1120, U.S.A.}\\  {\tt
paniak@umich.edu}
\\[6mm]

{\bf R.J. Szabo}
\\[3mm]
{\it Department of Mathematics\\ Heriot-Watt University\\ Scott
  Russell Building, Riccarton,
Edinburgh EH14 4AS, U.K.}\\  {\tt R.J.Szabo@ma.hw.ac.uk}
\\[30mm]

\end{center}

\begin{abstract}

\baselineskip=12pt

The correlation functions of open Wilson line operators in
two-dimensional Yang-Mills theory on the noncommutative torus are
computed exactly. The correlators are expressed in two equivalent
forms. An instanton expansion involves only topological numbers of Heisenberg
modules and enables extraction of the weak-coupling limit of the gauge
theory. A dual algebraic expansion involves only group theoretic
quantities, winding numbers and translational zero modes, and enables
analysis of the strong-coupling limit of the gauge theory and the
high-momentum behaviour of open Wilson lines. The dual expressions can be
interpreted physically as exact sums over contributions from virtual
electric dipole quanta.

\end{abstract}

\end{titlepage}


\newsection{Introduction and Summary}

Field theories on noncommutative spaces continue to challenge
the conventional wisdom of quantum field theory
(see~\cite{ks}--\cite{sz1} for reviews and exhaustive lists of
references). They can be naturally embedded into string theory and as
such possess many unusual non-local effects which are not observed in
their commutative counterparts. Striking differences between these
models and ordinary quantum field theories are readily seen in
perturbation theory. The perturbation series exhibits poles in the
noncommutativity parameter $\theta$ at each finite order, and mixes
ultraviolet and infrared modes in a way that appears to make
conventional renormalization schemes inapplicable to this class of
field theories. However, such non-analyticities and instabilities may
simply be an artifact of the perturbative approximation. In order to truly
understand how noncommutative field theories differ from ordinary
ones, and what they can teach us about string dynamics in background
fields, it is important to have a good understanding of the
non-perturbative physics of these models.

A good place to look for exactly solvable models is in two spacetime
dimensions. In particular, inspired by the wealth of methods that can be
used to solve ordinary gauge theory in two dimensions (see~\cite{cmr}
for a review), the gauge invariant correlation functions of
two-dimensional noncommutative Yang-Mills theory have been studied
from various different points of view with an eye towards
understanding generic features of its exact solution. Perturbative studies of
the analytic structure of Wilson loop correlators may be found
in~\cite{bnt}--\cite{torr}. Fluxon representations in the large $N$
limit of open Wilson line correlators are given
in~\cite{gsv}. The high-energy behaviour of open Wilson lines at large
$N$ is studied within the ladder approximation in~\cite{bv}. Numerical
simulations of both Wilson loops and open Wilson lines based on the
reduced model representation of noncommutative gauge theory may be
found in~\cite{Bietenholz:2002ch}--\cite{Bietenholz:2002ev}.

In this paper we shall build on the exact non-perturbative expression for the
vacuum amplitude of two-dimensional gauge theory on the
noncommutative torus that was obtained in~\cite{inprep}. We shall
present two main achievements. First of all, we will derive the exact
expression for the pair correlator of open Wilson line operators for
the first time. This will exhibit the strength of the technique
proposed in~\cite{inprep} at obtaining analytic results in gauge
theory on the noncommutative torus. Secondly, we will derive an
expansion dual to that of~\cite{inprep} which is analytic in the
Yang-Mills coupling constant. This series elucidates precisely how
noncommutativity alters gauge theory in two dimensions. In the
commutative case, the partition function and observables can be
evaluated by using heat kernel methods, leading to an explicit
expansion into irreducible representations of the gauge group which
may be parameterized by integer-valued Young tableaux row
variables. In the
noncommutative case, the resummed partition function shows that the
group theory row integers are promoted to doublets of integers which
generically cannot be decoupled. The open Wilson line calculations
then serve to show that the promoted group theory doublet can be
interpreted as consisting of a representation (winding) integer plus a
translational zero mode (momentum). This gives a very precise
demonstration of how noncommutativity interlaces colour and spacetime
degrees of freedom. These calculations also provide a basis for
potentially understanding how to make sense of the large $N$ limit of the
solution to gauge theory on the noncommutative torus, and hence a
possible realization of a string representation of two-dimensional
noncommutative Yang-Mills theory. Whether or not a sensible and
non-trivial large $N$ limit is possible on the noncommutative torus is
presently not clear, and this issue deserves further investigation.

\subsection{Outline and Summary of Results}

In the remainder of this section we will sketch the structure of the
rest of the paper and summarize our main findings. All facts used in
this paper concerning noncommutative gauge theory can be found
in~\cite{inprep} and the review articles~\cite{ks}--\cite{sz1}. In section~2
we
will start from the expansion, derived in~\cite{inprep}, of the
partition function of noncommutative Yang-Mills theory in terms of
partitions of the topological numbers characterizing a projective
module over the noncommutative torus. We will re-interpret this series
in terms of a dual description which involves the integer-valued
charges and momenta of collections of virtual electric dipoles, the
non-local quanta that characterize generic noncommutative field
theories~\cite{Sheikh1,BigSuss}. One advantage of this dual
representation is that with it the partition function is manifestly an
analytic function of the noncommutativity parameter, thus showing
explicitly that it is smooth in $\theta$. The simplifications which
occur in this expansion for rational values of the noncommutativity parameter
are investigated and the $SL(2,\zed)$ transformation properties of the
dual electric and momentum charges are derived. For irrational values
of the noncommutativity parameter such simplifications are not
possible, although a rational limiting process produces an expression
for the partition function which is reminiscent of a large $N$ matrix
model. Finally, we investigate an expansion in small values of
$\theta$ and show that noncommutative Yang-Mills theory is
asymptotically equivalent to generalized Yang-Mills
theory~\cite{genYM1}--\cite{genYM3} with exponentially small
corrections. It is important here that the series is only asymptotic,
so that the truncation to any finite order in the $\theta$-expansion
doesn't make sense, as expected from renormalizability arguments.

In section~3 we turn our attention to the study of non-local observables
in ordinary Yang-Mills theory on the torus. In particular, we consider the
pair correlator of simple Polyakov loops with integer winding number. Using
some well-known machinery, we derive an exact expression for these correlation
functions in terms of an instanton expansion. This represents the
first calculation of the Polyakov loop anti-loop correlator in the
instanton representation of commutative Yang-Mills theory. The Fourier
dual of this result produces a momentum-dependent loop correlator
which involves structures that are native to noncommutative geometry.

In section~4 we analyse the pair correlators of open Wilson lines in
the noncommutative case, which yield the interaction amplitudes
between pairs of electric dipoles. The correlation functions are
computed, via the arguments of~\cite{inprep}, exactly for all values of
the noncommutativity parameter $\theta$ by generalizing the momentum
space Polyakov loop pair correlator to noncommutative space using
Morita covariance under $SL(2,\zed)$ transformations of the
topological and dual charge-momentum integers. We illustrate how the
resulting correlation functions exhibit structures which appear in BPS
configurations of dipoles.

With an exact expression for the pair correlator at hand, in section~5 we
proceed to investigate its properties at both weak and strong
coupling. The weak-coupling limit of the correlator is shown to depend
on an $SL(2,\zed)$-invariant pairing of the topological and
charge-momentum integers in K-theory. Physically, the weak-coupling
results reveal new insights into commutative Yang-Mills theory with
correlation functions presented in a momentum representation. In
particular, it is shown that even the commutative theory naturally
depends on both charge and momentum quantum numbers which characterize
BPS states. We also compute the strong-coupling expansion of the
correlation function of open Wilson lines. This form gives the
clearest perspective on the dual nature of the topological and
charge-momentum integers, and it makes explicit the twisting together
of colour-electric and spacetime degrees of freedom in the
noncommutative gauge theory. Moreover, it can be used to show that the
correlator is a smooth function of $\theta$, thereby
suggesting that any non-analyticities in perturbative calculations are
resummable artifacts.

Finally, in section~6 we investigate the high-momentum dependence of
the correlators. We show that the correlation functions of open Wilson
line operators in two-dimensional noncommutative Yang-Mills theory are
uniformly bounded in the Yang-Mills coupling constant. In particular,
the correlator does not exhibit any exponential growth with momentum
like its four-dimensional counterpart
does~\cite{ghi}--\cite{DharKit1}. In fact, we establish explicitly that
the correlation functions vanish in the limit of large momentum.

\newsection{The Instanton Expansion and Group Theoretic
  Deformations\label{InstExp}}

The partition function of two-dimensional quantum gauge theory on the
noncommutative torus is given by the Euclidean functional integral
\beq
Z_{p,q}(A,\theta)=\int\DD{\cal A}~\exp\left[-\frac1{2g^2}\,\int\dd^2x~
\Tr\left(F-\frac{2\pi^2\,b_{p,q}}A\right)^2\,\right] \ ,
\label{Zpqfnint}\eeq
where $g$ is the Yang-Mills coupling constant and $A$ is the area of
  the square torus ${\bf T}^2$. Here $F=\partial_1{\cal
  A}_2-\partial_2{\cal A}_1+{\cal A}_1\star {\cal A}_2-{\cal
  A}_2\star{\cal A}_1$ is the noncommutative field strength
  tensor which is defined using the usual star-product with a
  dimensionless noncommutativity parameter $\theta$. The
  anti-Hermitian $U(N)$ gauge field $\cal A$ defines a connection of
  a fixed Heisenberg module over the noncommutative
  torus of topological numbers $(p,q)\in\zed^2$, dimension
  $p-q\,\theta>0$, and rank $N={\rm gcd}(p,q)$. In the
  commutative case $\theta=0$, this makes a distinction between
  physical Yang-Mills theory, which takes into account the sum over all
  magnetic fluxes~$q$ (i.e. all topological classes of principal
  $U(p)$ bundles over ${\bf T}^2$), and Yang-Mills theory defined on a
  fixed projective module. In the noncommutative case, only the latter
  gauge theory appears to be amenable to an unambiguous definition, as
  the topological numbers $(p,q)$ do not generically decouple in this case. We
  have also subtracted the constant curvature
\beq
b_{p,q}=\frac q{p-q\,\theta}
\label{bpq}\eeq
of the module in (\ref{Zpqfnint}) which corresponds to the global
  minimum of the noncommutative Yang-Mills action. It is a topological
  term whose only essential role is to ensure that the action is
  invariant under gauge Morita duality transformations.

The functional integral (\ref{Zpqfnint}) is given exactly by its
semi-classical
approximation, owing to a hidden supersymmetry present in the path integral
which kills higher quantum loop corrections. It can thereby be
represented as a sum over solutions of the classical noncommutative
Yang-Mills equations. This leads to the explicit expansion~\cite{inprep}
\bea
Z_{p,q}(A,\theta)&=&\sum_{\stackrel{\scriptstyle{\rm partitions}}
{\scriptstyle(\vp,\vq)}}\,\frac{(-1)^{|\vnu|}}{\prod_a\nu_a!}\,
\prod_{k=1}^{|\vnu|}\sqrt{\frac{2\pi^2}{g^2A\,
(p_k-q_k\,\theta)^3}}\non&&\times\,\exp\left[-\frac{2\pi^2}{g^2A}\,
\sum_{k=1}^{|\vnu|}\left(p_k-q_k\,\theta\right)
\left(\frac{q_k}{p_k-q_k\,\theta}-\frac q{p-q\,\theta}\right)^2\right] \ ,
\label{Zpqpartitionsum}\eeq
where the sum is over all ``partitions'' of the
topological numbers $(p,q)$, which are the collections of integers
$(\vp,\vq)=\{(p_k,q_k)\}$ obeying the constraints
\bea
p_k-q_k\,\theta&>&0 \ , \non\sum_k\left(p_k-q_k\,\theta\right)&=&
p-q\,\theta \ , \non\sum_kq_k&=&q \ .
\label{partitiondef}\eeq
The integer $\nu_a>0$ is the number of partition components
$(p_k,q_k)\in(\vp,\vq)$ which have the $a^{\rm th}$ least dimension
$p_a-q_a\,\theta$, while $|\vnu|=\nu_1+\nu_2+\dots$ is the total
number of components in the partition
$(\vp,\vq)$.\footnote{\baselineskip=12pt Notation: For a given
  partition, the integer $k=1,\dots,|\vnu|$ labels its components
  $(p_k,q_k)$, while $a$ labels the degeneracy integers $\nu_a$. Since
the coupling constant and area of the torus enter the partition function
only through the dimensionless combination $g^2A$, for brevity we
display explicitly only the area dependence of all observables.} When
$\theta=0$, the last constraint in (\ref{partitiondef}) distinguishes the
partition function (\ref{Zpqpartitionsum}) from that of physical
Yang-Mills theory. Physical commutative Yang-Mills theory is realized
as an appropriately weighted sum over sectors with distinct Chern
numbers $q$.

The set of all partitions for a given Heisenberg module are in
one-to-one correspondence with classical gauge field
configurations. The exponential factor in (\ref{Zpqpartitionsum})
is the classical contribution to the Yang-Mills partition function
(\ref{Zpqfnint}). By resumming (\ref{Zpqpartitionsum}) over gauge inequivalent
partitions~\cite{inprep}, we may thereby express it as a sum over unstable
instantons. The exponential prefactors in (\ref{Zpqpartitionsum}) then
represent the {\it exact} corrections due to quantum fluctuations
around a classical solution and are determined by a finite, non-trivial
perturbative expansion about each instanton.

\subsection{The Dual Expansion\label{DualExp}}

The expansion (\ref{Zpqpartitionsum}) is inherently non-perturbative
as it contains terms of order $\e^{-1/g^2A}$. It is natural to ask
if there exists a corresponding ``dual'' expansion which is
analytic in the Yang-Mills coupling constant. For this, we will drop
the background flux $b_{p,q}$ in (\ref{Zpqfnint}) for simplicity and
rewrite the partition sum (\ref{Zpqpartitionsum}) as series over {\it
  unconstrained} integers as
\bea
Z_{p,q}(A,\theta)&=&\sum_{\vp,\vq}\frac{(-1)^{|\vnu|}}{\prod_a\nu_a!}~
\delta^{~}_{p\,,\,\sum_kp_k}~\delta^{~}_{q\,,\,\sum_kq_k}\non&&\times\,
\prod_{k=1}^{|\vnu|}\sqrt{\frac{2\pi^2}{g^2A}}~
\int\limits_{0^+}^\infty\frac{\dd z_k}
{(z_k)^{3/2}}~\delta(z_k-p_k+q_k\,\theta)~\e^{-\frac{2\pi^2}{g^2A}\,
\frac{(q_k)^2}{z_k}} \ .
\label{Zpqunconstr}\eea
The integrations over $z_k>0$ enforce the dimension positivity
constraints in (\ref{partitiondef}), while the Kronecker
delta-functions enforce the last two constraints. By resolving the
delta-functions over the topological charges $q_k$ and the dimension
variables $z_k$ we get
\bea
Z_{p,q}(A,\theta)&=&\sum_{\vp,\vq}\frac{(-1)^{|\vnu|}}{\prod_a\nu_a!}~
\delta^{~}_{p\,,\,\sum_kp_k}~\int\limits_0^1\dd\lambda~
\e^{-2\pi\ii\lambda\,q}\non&&\times\,\prod_{k=1}^{|\vnu|}
\sqrt{\frac{2\pi^2}{g^2A}}~\int\limits_{0^+}^\infty\frac{\dd
z_k}{(z_k)^{3/2}}~
\int\limits_{-\infty}^\infty\frac{\dd x_k}{2\pi}~
\e^{\ii x_k(z_k-p_k+q_k\,\theta)}~\e^{2\pi\ii\lambda\,q_k-
\frac{2\pi^2}{g^2A}\,\frac{(q_k)^2}{z_k}} \ . \non&&
\label{Zpqdeltaresolve}\eea
The $q_k$ series are Gaussian sums, and so they can each be
rewritten by using the Poisson resummation formula
\beq
\sum_{m=-\infty}^\infty\e^{-\pi hm^2-2\pi\ii b\,m}=\frac1{\sqrt h}
{}~\sum_{q=-\infty}^\infty\e^{-\pi(q-b)^2/h} \ .
\label{Poisson}\eeq
Then the resulting integrations over $x_k$ are also Gaussian, and we
find
\bea
Z_{p,q}(A,\theta)&=&\sum_{\vp,\vm}\frac{(-1)^{|\vnu|}}{\prod_a\nu_a!}~
\delta^{~}_{p\,,\,\sum_kp_k}~\int\limits_0^1\dd\lambda~
\e^{-2\pi\ii\lambda\,q}\non&&\times\,\prod_{k=1}^{|\vnu|}
\sqrt{\frac{2\pi^2}{g^2A\,\theta^2}}~
\int\limits_{0^+}^\infty\frac{\dd z_k}{(z_k)^{3/2}}~
\e^{-\frac{g^2A}2\,z_k(m_k+\lambda)^2}\non&&\times\,
\exp\left[-\frac{2\pi^2}{g^2A\,\theta^2\,z_k}\,\left(p_k-z_k-\ii
\,\frac{g^2A\,\theta\,z_k}{2\pi}\,\Bigl(\lambda+m_k\Bigr)\right)^2
\,\right] \ .
\label{Zpqqtom}\eea

Now we repeat the same procedure for the sum over the $p_k$'s in
(\ref{Zpqqtom}). Fourier resolving the constraints on the $p_k$'s by a
circular coordinate $\mu\in[0,1]$ leaves a Gaussian sum in each $p_k$,
which may be Poisson resummed using (\ref{Poisson}) in terms
of dual integers $n_k$. This gives
\bea
Z_{p,q}(A,\theta)&=&\sum_{\vn,\vm}\frac{(-1)^{|\vnu|}}{\prod_a\nu_a!}\,
\int\limits_0^1\dd\mu~\e^{-2\pi\ii\mu\,p}\,\int\limits_0^1\dd\lambda~
\e^{-2\pi\ii\lambda\,q}\non&&\times\,\prod_{k=1}^{|\vnu|}~
\int\limits_{0^+}^\infty\frac{\dd z_k}{z_k}~\e^{-\frac{g^2A}2\,
z_k\bigl(m_k+\lambda+\theta\,(\mu-n_k)\bigr)^2}~
\e^{2\pi\ii z_k(\mu-n_k)} \ .
\label{Zpqpton}\eea
We now shift the circular coordinate $\lambda\to\lambda-\theta\,\mu$
in order to eliminate the $\mu$-dependence of the quadratic exponent in
(\ref{Zpqpton}). Then the shifted integration range is
$\lambda\in[\theta\,\mu\,,\,1+\theta\,\mu]$. But the integrand of
(\ref{Zpqpton}) is easily seen to be a periodic function of $\lambda$
with period~1. As a consequence, since the $\lambda$ integral goes
over a single period, it is independent of the
offset~$\theta\,\mu$. In this way we arrive at our final expression
for the resummed, dual partition function as
\bea
Z_{p,q}(A,\theta)&=&\sum_{\vn,\vm}\frac{(-1)^{|\vnu|}}{\prod_a\nu_a!}\,
\int\limits_0^1\dd\mu~\e^{-2\pi\ii\mu\,(p-q\,\theta)}\,
\int\limits_0^1\dd\lambda~
\e^{-2\pi\ii\lambda\,q}\non&&\times\,\prod_{k=1}^{|\vnu|}~
\int\limits_{0^+}^\infty\frac{\dd z_k}{z_k}~\e^{-\frac{g^2A}2\,
z_k(m_k-n_k\,\theta+\lambda)^2}~\e^{2\pi\ii z_k(\mu-n_k)} \ .
\label{Zpqresumfinal}\eea
Note that in this expansion the multiplicities $\nu_a$ are associated
with the distinct numbers $m_a-n_a\,\theta$.

This dual form makes manifest the partition function as an analytic
function of the noncommutativity parameter $\theta$, as was proven
in~\cite{inprep} directly from the original instanton expansion
(\ref{Zpqpartitionsum}). Furthermore, while the instanton form
(\ref{Zpqpartitionsum}) is useful for extracting the weak coupling
limit of the gauge theory~\cite{inprep}, the resummed version
(\ref{Zpqresumfinal}) makes tractable the large area (strong coupling) limit,
from which we may conclude that the partition function on the noncommutative
plane is a constant independent of all coupling parameters. As we will see
below, it also allows one to make some explicit substantial simplifications of
$Z_{p,q}(A,\theta)$ for rational values of $\theta$. In this way the dual
expansion (\ref{Zpqresumfinal}) will give a very precise description of
how rational noncommutative gauge theories relate to commutative
ones.

We will see later on that the quantities $m_k-n_k\,\theta$ and $n_k$ in
(\ref{Zpqresumfinal}) correspond to electric flux and momentum, respectively,
carried by the elementary quanta of the noncommutative gauge theory. The dual
expansion thereby rewrites the vacuum energy in terms of contributions from
virtual electric dipoles, with the quadratic terms in the exponential
representing their zero-mode kinetic energy. The Poisson duality used above
may
therefore be regarded as an electric-magnetic duality which exchanges
instantons and dipoles. In this section we shall focus on the group
theoretical
aspects of the strong-coupling expansion (\ref{Zpqresumfinal}).

\subsection{Group Theory at $\theta=0$\label{GroupTheory}}

To get a feel for what the dual instanton expansion
(\ref{Zpqresumfinal}) represents, let us briefly consider the commutative case
$\theta=0$ and recall the well-known exact solution in this case in
terms of group theoretical techniques~\cite{cmr}. We start from {\it ordinary}
$U(p)$ Yang-Mills theory on a finite cylinder within the Hamiltonian
formalism, whose Hilbert space of physical states is the space of
$L^2$-class functions with respect to the normalized, invariant Haar measure
$[\dd U]$ on the unitary group $U(p)$. A basis for this Hilbert space is
provided by the characters
\beq
\chi^{~}_R(U)=\Tr^{~}_R~U
\label{chiRU}\eeq
in the irreducible unitary representations $R$ of $U(p)$, where $U$ is
the holonomy of the gauge field around the cycle of the cylinder at some fixed
time slice. The Hamiltonian is diagonalized in this representation basis and
is essentially the Laplacian on the group manifold of $U(p)$. Its
eigenvalue on the wavefunction (\ref{chiRU}) is proportional to the second
Casimir invariant $C_2(R)$ of the representation $R$.

{}From these facts one can immediately write down the cylinder amplitude
corresponding to propagation between two states characterized by holonomies
$U_1$ and $U_2$ as
\beq
K_p(A;U_1,U_2)=\sum_R\chi_R^{~}(U_1)\,\chi_R^*(U_2)~
\e^{-\frac{g^2A}2\,C_2(R)} \ ,
\label{cylinderampl}\eeq
where $A$ is the area of the cylinder. This is just the standard heat
kernel on the $U(p)$ group. We can now glue the two ends of the
cylinder together by setting $U_1=U_2=U$ and integrate over $U$
using the fact that characters are orthonormal with respect to integration
over
the Haar measure of $U(p)$,
\beq
\int\limits_{U(p)}[\dd U]~\chi_R^{~}(U)\,\chi^*_{R'}(U)=
\delta_{R,R'} \ .
\label{charortho}\eeq
This gives the partition function for Yang-Mills theory on a torus
${\bf T}^2$ of area $A$ as the vacuum amplitude
\beq
Z_p(A)=\int\limits_{U(p)}[\dd U]~K_p(A;U,U)=
\sum_R\e^{-\frac{g^2A}2\,C_2(R)} \ .
\label{Zptorus}\eeq

One can make the sum in (\ref{Zptorus}) explicit by noting that
the irreducible representations of the $U(p)$ structure group are
parameterized by decreasing sets $\vm=(m_1,\dots,m_p)$ of $p$ integers
\beq
+\infty>m_1>m_2>\dots>m_p>-\infty
\label{irrepms}\eeq
which are shifted highest weights that determine the lengths of the
rows of the corresponding Young tableaux. Up to an irrelevant
constant, the quadratic Casimir is given in terms of these row integers as
\beq
C_2(R)=C_2(\vm)=\sum_{a=1}^p\left(m_a-\frac{p-1}2\right)^2
\label{C2Rn}\eeq
and is symmetric under permutations of the $m_a$'s. One can thereby
write the sum in (\ref{Zptorus}) as a sum over all non-coincident
integers $m_1\neq m_2\neq\cdots\neq m_p$, which can then be turned
into a sum over {\it all} $\vm\in\zed^p$ by inserting
\beq
\det_{1\leq a,b\leq p}\,\left(\delta_{m_a,m_b}\right)=
\sum_{\sigma\in S_p}(-1)^{|\sigma|}\,\prod_{a=1}^p
\delta_{m_a,m_{\sigma(a)}} \ .
\label{deltainsert}\eeq
The permutation symmetry of (\ref{C2Rn}) then implies that
(\ref{deltainsert}) truncates to a sum over conjugacy classes
$[1^{\nu_1}\,2^{\nu_2}\cdots p^{\nu_p}]$ of the symmetric group $S_p$
which are labelled by partitions $\vnu=(\nu_1,\dots,\nu_p)$ of $p$,
with $\nu_a\geq0$ the number of elementary cycles of length $a$ in
$[1^{\nu_1}\,2^{\nu_2}\cdots p^{\nu_p}]$ satisfying
\beq
\nu_1+2\,\nu_2+\dots+p\,\nu_p=p \ .
\label{nupartition}\eeq
In this way, by dropping some irrelevant factors, (\ref{Zptorus})
becomes
\beq
Z_p(A)=\sum_{\vm\in\zeds^p}~\sum_{\stackrel{\scriptstyle{\rm partitions}}
{\scriptstyle\vnu}}~\prod_{a=1}^p\frac{(-1)^{\nu_a}}{a^{\nu_a}\,\nu_a!}~
\e^{-\frac{g^2A}2\,a\,C_2\bigl(m_{\nu_1+\dots+\nu_{a-1}+1}\,,\,\dots\,,\,
m_{\nu_1+\dots+\nu_a}\bigr)} \ ,
\label{Zpcycle}\eeq
where the exponential prefactor comes from the sign and order of the
conjugacy class $[1^{\nu_1}\,2^{\nu_2}\cdots p^{\nu_p}]$, and we have defined
$\nu_0\equiv0$.

After repeated applications of the Poisson resummation formula
(\ref{Poisson}), we arrive at the expansion (\ref{Zpqpartitionsum}) in the
commutative case $\theta=0$, with a weighted sum over all topological
charges $q$,
\beq
Z_p(A)=\sum_{q=-\infty}^\infty\e^{\pi\ii(p-1)\,q}~Z_{p,q}(A,0) \ .
\label{Zpcomm}\eeq
The details of this computation may be found in~\cite{inprep}. Note
that at the level of the expansion (\ref{Zpcycle}), the definition of
partition is slightly different from that given at the beginning of
this section, because the magnetic charge $q$ does not enter explicitly
into the physical partition function. Namely, here the partitions are
the collections of integers $\vp=\{p_k\}$, $0<p_k\leq p$ associated with the
conjugacy classes of the symmetric group $S_p$, each component of
$\vp$ specifying the length of a cycle in a class. The first two
constraints in (\ref{partitiondef}) for $\theta=0$ are still met, but
now we allow for vanishing $\nu_a$, the number of components in $\vp$
of magnitude~$a$.

Let us now return to (\ref{Zpqresumfinal}) in light of this analysis. In the
limit of vanishing noncommutativity parameter, the sums over $m_k$ and $n_k$
in
(\ref{Zpqresumfinal}) decouple. The series over $n_k$ yields a
periodic delta-function
\beq
\sum_{n_k=-\infty}^\infty\e^{2\pi\ii z_kn_k}=\sum_{p_k=1}^\infty
\delta(z_k-p_k) \ ,
\label{nkseriescomm}\eeq
where we have used the fact that $z_k>0$. Then the integration over $\mu$ in
(\ref{Zpqresumfinal}) imposes the constraint $\sum_kp_k=p$, i.e. that
$\vp\in\zed_+^p$ form a partition of the rank $p$ of the structure group. The
sign factors in (\ref{Zpqresumfinal}) ensure that no pairs of integers $m_k$
coincide, and the integers $p_k$ enumerate coincident values of the $m_k$. By
using the symmetry of (\ref{Zpqresumfinal}) under arbitrary permutations of
the
$m_k$, summing over the $p_k$ thereby produces the dual expansion of the
commutative partition function as
\beq
Z_{p,q}(A,0)=\int\limits_0^1\dd\lambda~\e^{-2\pi\ii\lambda\,q}~
\sum_{m_1>\dots>m_p}\e^{-\frac{g^2A}2\,\sum_a(m_a+\lambda)^2} \ .
\label{Zpqdualcomm}\eeq
The weighted sum (\ref{Zpcomm}) of (\ref{Zpqdualcomm}) reproduces the
physical Yang-Mills partition function (\ref{Zptorus}) as a sum
over the Young tableaux weights (\ref{irrepms}) with Casimir invariant
(\ref{C2Rn}). Generally, (\ref{Zpqdualcomm}) represents the heat kernel
expansion of commutative Yang-Mills theory on a torus with the integration
enforcing the constraints, within an algebraic setting, on the topological
numbers of the given projective module.

\subsection{Gauge Morita Equivalence at Rational $\theta$\label{MoritaPart}}

For integer values $\theta=r\in\zed$ of the noncommutativity parameter, we can
eliminate the $n_k$ dependence of the quadratic exponent in
(\ref{Zpqresumfinal}) by shifting $m_k$ by the integer $r\,n_k$, and in the
same manner as in the previous subsection recover the heat kernel expansion of
commutative $U(p-rq)$ gauge theory. In this way we give a
direct proof, at a fully non-perturbative level, of the anticipated
equivalence
between noncommutative Yang-Mills theory in the limit $\theta\to\infty$ and
the planar limit of ordinary non-Abelian gauge theory. This also shows
explicitly that we may restrict attention to noncommutativity parameters in
the
range $\theta\in[0,1)$, as the integer part of $\theta$ can always be shifted
away.

For rational values $\theta=r/s$, with $r<s$ relatively prime positive
integers, one can simplify (\ref{Zpqresumfinal}) to a finite weighted sum over
contributions from commutative Yang-Mills theory.
For this, we consider separately the subseries $n_k=s\,d_k+a_k$, where
$d_k\in\zed$ and $a_k=0,1,\dots,s-1$. The partition function
(\ref{Zpqresumfinal}) is then given by the sum over $d_k$ and $a_k$, and again
by shifting $m_k$ by the integer $r\,d_k$ we can decouple the sums over $d_k$
to get
\bea
Z_{p,q}(A,r/s)&=&\sum_{\vd,\vm}~\sum_{a_1,a_2,\dots\,=0}^{s-1}~
\frac{(-1)^{|\vnu|}}{\prod_a\nu_a!}\,\int\limits_0^1\dd\mu~
\e^{-2\pi\ii\mu\,(p-r\,q/s)}\,\int\limits_0^1\dd\lambda~
\e^{-2\pi\ii\lambda\,q}\non&&\times\,\prod_{k=1}^{|\vnu|}~
\int\limits_{0^+}^\infty\frac{\dd z_k}{z_k}~
\e^{-\frac{g^2A}2\,z_k(m_k-r\,a_k/s+\lambda)^2}~
\e^{2\pi\ii z_k(\mu-s\,d_k-a_k)} \ .
\label{dkdecouple}\eeq
The series over $d_k$ and integral over $z_k$ force $s\,z_k$ to be a positive
integer $p_k'$, giving
\bea
Z_{p,q}(A,r/s)&=&\sum_{\vp'\,:~p_k'>0}~\sum_\vm~
\sum_{a_1,a_2,\dots\,=0}^{s-1}~
\frac{(-1)^{|\vnu|}}{\prod_a\nu_a!}\,\int\limits_0^1\dd\mu~
\e^{-2\pi\ii\mu\,(p-r\,q/s)}\,\int\limits_0^1\dd\lambda~
\e^{-2\pi\ii\lambda\,q}\non&&\times\,\prod_{k=1}^{|\vnu|}\,
\frac s{p_k'}~\e^{-\frac{g^2A}{2s}\,p_k'(m_k-r\,a_k/s+\lambda)^2}~
\e^{2\pi\ii p_k'(\mu-a_k)/s} \ .
\label{pkprimesum}\eea
After rescaling $\mu\to s\,\mu$, the integrand of (\ref{pkprimesum}) is
invariant under integer shifts of the circular coordinate $\mu$. Averaging
(\ref{pkprimesum}) over the integer shifts in $\mu$ from 0 to $s-1$ then
yields a manifestly $\zed$-periodic expression in $\mu$ as
\bea
Z_{p,q}(A,r/s)&=&\sum_{\vp'\,:~p_k'>0}~\sum_\vm~
\sum_{a_1,a_2,\dots\,=0}^{s-1}~
\frac{(-1)^{|\vnu|}}{\prod_a\nu_a!}\,\int\limits_0^1\dd\mu~
\e^{-2\pi\ii\mu\,(s\,p-r\,q)}\,\int\limits_0^1\dd\lambda~
\e^{-2\pi\ii\lambda\,q}\non&&\times\,\prod_{k=1}^{|\vnu|}\,
\frac s{p_k'}~\e^{-\frac{g^2A}{2s}\,p_k'(m_k-r\,a_k/s+\lambda)^2}~
\e^{2\pi\ii p_k'(\mu-a_k/s)} \ .
\label{zedperiodicsum}\eea

It is now clear from (\ref{zedperiodicsum}) that the integration over $\mu$
enforces the constraint
\beq
\sum_kp_k'=s\,p-r\,q\equiv p' \ .
\label{sumkpkprime}\eeq
As in the previous subsection, we will express this partitioning of the
integer $p'$, along with the sign factors in (\ref{zedperiodicsum}), with
$p_k'$ enumerating coincident values of the sets of integers $m_k$ and $a_k$.
With this, the partition function (\ref{zedperiodicsum}) can be written in
terms of $2p'$ non-coincident integers as
\beq
Z_{p,q}(A,r/s)=\sum_{\stackrel{\scriptstyle\va\in\zeds_s^{p'}}
{\scriptstyle a_b\neq a_c}}\e^{-\frac{2\pi\ii}s\,\sum_ba_b}~
\int\limits_0^1\dd\lambda~\e^{-2\pi\ii\lambda\,q}~
\sum_{\stackrel{\stackrel{\scriptstyle\vm\in\zeds^{p'}}
{\scriptstyle{~~}}}{\scriptstyle m_b>m_c}}
\e^{-\frac{g^2A}{2s}\,\sum_b(m_b-r\,a_b/s+\lambda)^2} \ .
\label{Zmanoncoinc}\eeq
The particularly interesting aspect of the expansion (\ref{Zmanoncoinc}) is
that, upon comparison with the heat kernel expansion (\ref{Zpqdualcomm}), it
shows that the Young tableaux row integers $m_b$ are replaced in the
noncommutative case by the fractional, $\frac1s$-valued variables
$m_b+r\,a_b/s$. The momentum integers $a_b$ also appear in this case to
parameterize
states of some $\zed_s$-valued theta-sector of the two-dimensional gauge
theory, as would arise from an $SU(s)$ structure group. This illustrates how
noncommutativity deforms group representation quantities, similarly to the way
in which it alters the dimensions of projective modules over the torus.

We can take this description of ``noncommutative group theory'' even further
by exploiting Morita equivalence with commutative Yang-Mills theory in
this case. The discrete M\"obius transformation which maps the
noncommutativity parameter $\theta=r/s$ to $\theta'=0$ is
parameterized by the $SL(2,\zed)$ matrix
\beq
\begin{pmatrix}\,s&-r\,\cr\,r'&s'\,\cr\end{pmatrix} \ , ~~
r\,r'+s\,s'=1 \ .
\label{SL2Zcommtransf}\eeq
The topological numbers associated to any partition of a given Heisenberg
module transform as $SL(2,\zed)$ doublets under the gauge extension of Morita
equivalence,
\beq
\begin{pmatrix}\,\vp'\,\cr\,\vq'\,\cr\end{pmatrix}=
\begin{pmatrix}\,s&-r\,\cr\,r'&s'\,\cr\end{pmatrix}
\begin{pmatrix}\,\vp\,\cr\,\vq\,\cr\end{pmatrix} \ ,
\label{pqSL2Zdoublets}\eeq
so that by Poisson duality we also have
\beq
\begin{pmatrix}\,\vm'\,\cr\,\vn'\,\cr\end{pmatrix}=
\begin{pmatrix}\,s&-r\,\cr\,r'&s'\,\cr\end{pmatrix}
\begin{pmatrix}\,\vm\,\cr\,\vn\,\cr\end{pmatrix} \ .
\label{nmSL2Zdoublets}\eeq
This gives a definition of the action of Morita duality on the
algebraic numbers of the noncommutative gauge theory, rather than the
conventional definition in terms of topological numbers. In addition,
the dimension variables and the dimensionless Yang-Mills coupling constant
transform in this particular case as
\bea
z_k'&=&s\,z_k \ , \non\left(\,g^2A\right)'&=&\frac{g^2A}{s^3} \ .
\label{dimcouplingtransf}\eea

With these transformation rules, we can obtain a concise proof of
the relation between the partition function (\ref{Zmanoncoinc}) at
rational $\theta$ and a commutative Yang-Mills theory.
Using the transformation rules (\ref{nmSL2Zdoublets})
of $SL(2,\zed)$ doublets  of dual topological numbers $(m_b,a_b)$,
along with (\ref{dimcouplingtransf}), we
obtain from (\ref{Zmanoncoinc}) the Morita dual partition function
\beq
Z_{p,q}(A,r/s)= \sum_{m_1>\dots>m_{p'}}~
\e^{\frac{2\pi\ii r'}s\,\sum_b m'_b}~
\int\limits_0^1\dd\lambda~\e^{-2\pi\ii\lambda\,q}~
\e^{-\frac{(g^2A)'}{2}\,\sum_b(m'_b + s\,\lambda)^2} \ .
\label{ZMoritadual}
\eeq
By rescaling $\lambda=\lambda'/s$, taking advantage of the
periodicity of the integrand and using the transformation rule for the
topological numbers (\ref{pqSL2Zdoublets}), we may bring
(\ref{ZMoritadual}) into the form
\beq
Z_{p,q}(A,r/s)=
\int\limits_0^1\dd\lambda'~\e^{-2\pi\ii\lambda'\,q' }~\sum_{m_1>\dots>m_{p'}}
\e^{-\frac{(g^2A)'}{2}\,\sum_b(m'_b +  \lambda')^2}~
\e^{\frac{2\pi\ii r'}s\, \sum_b (m'_b + \lambda')} \ .
\label{ZMoritadualfinalx}
\eeq
Comparing with the commutative partition function (\ref{Zpqdualcomm}), we
find that (\ref{ZMoritadualfinalx}) differs by an exponential factor which
indicates the presence of a background flux. The effect of this flux can
be made more explicit by absorbing it into quadratic terms in the energy
and by again shifting $\lambda'$.  The final result is
\beq
Z_{p,q}(A,r/s)= \e^{- \frac{2 \pi^2}{(g^2 A)'}\,\frac{ r'}{s^2}
\,(r'p' - 2 s\,q')}\,
\int\limits_0^1\dd\lambda'~\e^{-2\pi\ii\lambda'\,q' }~\sum_{m_1>\dots>m_{p'}}
\e^{-\frac{(g^2A)'}{2}\,\sum_b(m'_b +  \lambda')^2}  \ .
\label{ZMoritadualfinal}
\eeq
This form differs from the commutative result (\ref{Zpqdualcomm})
only by a global exponential prefactor which
is cancelled by the dual of the background flux $b_{p,q}$ in
(\ref{Zpqfnint}) that has been dropped from the present formulas.
In this way we have rederived the well-known  duality
between noncommutative Yang-Mills theory with coupling $g^2A$,
topological charge $q$ and rational-valued noncommutativity parameter
$\theta=r/s$, and ordinary non-Abelian gauge theory with coupling
constant $g^2A/s^3$ and structure group $U(sp-rq)$.

\subsection{Higher Casimir Operators at Irrational $\theta$\label{HigherCas}}

For irrational values of the noncommutativity parameter $\theta$, it
is not possible to reduce the dual partition indices appearing in the
quadratic form of (\ref{Zpqresumfinal}) to a single integer, as the electric
flux variables $m_k-n_k\,\theta$ cannot be decoupled in this case. The
partition function is always intrinsically a sum over the two-dimensional
dual lattice $\zed^2$ of K-theory charges. In the commutative and rational
cases the K-theory group ${\rm K}_0$ is isomorphic to
$\zed\oplus\zed$. Hence the charges are independent and may be
decoupled. In the irrational case the K-theory is instead naturally
isomorphic to the group $\zed+\zed\,\theta\subset\real$, and thus the
fluxes are topologically tied together. As we described in the
previous subsection, we may consider the effect of noncommutativity as
deforming Young tableaux row integers $m_k$ to non-integral ones
$m_k-n_k\,\theta$. As we will discuss further later on, this property is a
direct group theoretical reflection of the known mathematical property
that there is no well-defined notion of a structure group in
noncommutative gauge theory, because of the mixing between spacetime
and colour degrees of freedom through noncommutative gauge
transformations. Indeed, this is the main reason for the lack of a
definition of ``physical'' Yang-Mills theory on the noncommutative torus.

The group theoretic form (\ref{Zmanoncoinc}) of the rational partition
function illustrates this point in an interesting way. Let us consider
the irrational number $\theta$ as the limit of large positive integers $r$ and
$s$ with $\theta=\lim_{r,s}r/s$ fixed. The exponential sums in
(\ref{Zmanoncoinc}) over $b$ essentially run from 0 to
$s\,(p-q\,\theta)$ and are each weighted with a factor of $1/s$. In
the limit $s\to\infty$, these sums turn into integrals and the
partition function (\ref{Zmanoncoinc}) can be written as a path
integral
\bea
Z_{p,q}(A,\theta)&=&\prod_x~\int\limits_{\stackrel{\scriptstyle
m(x)\in\zeds}{\scriptstyle m(x)\neq m(y)}}\dd m(x)~
\int\limits_{\stackrel{\scriptstyle a(x)\in\zeds_+}
{\scriptstyle a(x)\neq a(y)}}\dd a(x)~\int\limits_0^1\dd\lambda~
\e^{-2\pi\ii\lambda\,q}\non&&\times\,\exp\left[-2\pi\ii
\int\limits_0^{p-q\,\theta}\dd x~a(x)-\frac{g^2A}2\,\int
\limits_0^{p-q\,\theta}\dd x~\Bigl(m(x)-\theta\,a(x)+\lambda
\Bigr)^2\,\right] \ . \non&&
\label{limZpathint}\eea
Note that the fields in (\ref{limZpathint}) are integer-valued and obey an
``exclusion'' principle which ensures that there are no continuous fields in
this mechanical system. This model defines a constrained but otherwise
Gaussian system, which resembles very much that which is obtained as
the large $N$ limit of Young tableaux in the planar limit of the heat
kernel representation of ordinary Yang-Mills theory~\cite{DougKaz}. Again this
illustrates the deformation of the group
theoretic representation by noncommutativity.

To understand this latter point better, let us return to the intermediate
form (\ref{Zpqqtom}) of the resummed partition function. Formally
performing the integrals over the dimension variables $z_k$ yields the
statistical sum
\bea
Z_{p,q}(A,\theta)'&=&{\sum_{\vp,\vm}}'~\frac{(-1)^{|\vnu|}}{\prod_a\nu_a!}~
\delta_{p\,,\,\sum_kp_k}^{~}~\int\limits_0^1\dd\lambda~
\e^{-2\pi\ii\lambda\,q}\non&&\times\,\prod_{k=1}^{|\vnu|}\,
\frac{\sqrt{\pi}}{|p_k|}\,\exp\left\{-\frac2{g^2A\,\theta^2}\,
\left[2\pi\,|p_k|\,\sqrt{\pi^2+\pi\ii g^2A\,\theta\,(m_k+\lambda)}
\right.\right.\non&&-\left.\left. p_k\,\Bigl(2\pi^2+\pi\ii
g^2A\,\theta\,(m_k+\lambda)\Bigr)\right] \right\} \ .
\label{Zpqzkint}\eea
The result (\ref{Zpqzkint}) omits from the sum those modules with
vanishing $p_k$. The full result is not so straightforward to express
in a simple way. However, this does not affect the expansion of
(\ref{Zpqzkint}) in the limit $g^2A\,\theta\to0$, to which we now
turn. In that limit, for positive $p_k$ the exponential prefactor is
identical to that of commutative gauge theory,
while the contributions from partitions with $p_k \leq 0$ are
exponentially suppressed. To leading orders in $g^2A\,\theta$, we
may then write the full partition function as
\bea
&&\lim_{g^2A\,\theta\to0}\,Z_{p,q}(A,\theta)~=~
\sum_{\stackrel{\scriptstyle\vp\in\zeds_+^p}{\scriptstyle
\sum_kp_k=p}}~\sum_{\vm\in\zeds^p}\,\frac{(-1)^{|\vnu|}}
{\prod_a\nu_a!}\,\int\limits_0^1\dd\lambda~\e^{-2\pi\ii\lambda\,q}
{}~\prod_{k=1}^{|\vnu|}\,\frac{\sqrt{\pi}}{p_k}\non&&~~~~~~~~~~\times\,
\exp\left\{-\frac{g^2A}2\,p_k\left[\left(m_k+\lambda\right)^2-\ii
\frac{g^2A\,\theta}{2\pi}\,\Bigl(m_k+\lambda\Bigr)^3
-\frac{5\left(\,g^2A\,\theta\right)^2}{16\pi^2}\,
\Bigl(m_k+\lambda\Bigr)^4\right.\right.\non&&~~~~~~~~~~+\left.\Biggl.
O\Bigl(\left(\,g^2A\,\theta\right)^3\Bigr)
\Biggr]\right\}+O\left(\e^{-1/g^2A\,\theta^2}\right) \ .
\label{Zpqzkintexpand}\eea

The expansion (\ref{Zpqzkintexpand}) gives a nice interpretation
of the noncommutative gauge theory as a particular modification of
ordinary Yang-Mills theory in the limit of small $g^2A\,\theta$.
Notice first of all that the corrections to the commutative
constraint $p_k>0$ are non-perturbative, exactly as is expected
from the instanton representation. This gives a
very clear picture of the dangers in viewing the theory as a
perturbative expansion in the noncommutativity parameter $\theta$. The
$m_k$'s to this order then have the usual interpretation as Young
tableaux row integers for $U(p)$, while the sums over the $p_k$'s can
be interpreted, as before, as a sum
over conjugacy classes enforcing the non-coincidence of the $m_k$'s.
Rewriting (\ref{Zpqzkintexpand}) with this restriction and dropping
irrelevant constants leads to a transparent form of the perturbative
corrections to commutative Yang-Mills theory as
\bea
\lim_{g^2A\,\theta\to0}\,
Z_{p,q}(A,\theta)&=&\int\limits_0^1\dd\lambda~\e^{-2\pi\ii\lambda\,q}\,
\sum_{m_1>\cdots>m_p}\exp\left\{-\frac{g^2A}2\,\sum_{a=1}^p
\biggl[\left(m_a+\lambda\right)^2\biggr.\right.\non &&
-\left.\left.\sum_{\ell=1}^\infty\,\frac{8\ii^\ell\,\left(1-\frac12\right)
\left(2-\frac12\right)\cdots\left(\ell-\frac12\right)}{\pi^\ell\,(\ell+2)!}
\,\left(\,g^2A\,\theta\right)^\ell\,\left(m_a+\lambda\right)^{\ell+2}
\right]\right\} \non&&+\,O\left(\e^{-1/g^2A\,\theta^2}\right) \ .
\label{Zpqzkintexpandfinal}\eea

The argument of the exponential in (\ref{Zpqzkintexpandfinal}) can
be interpreted in terms of Casimir operators. Generally, the
$\ell^{\,\rm th}$ Casimir operator eigenvalue in the representation $R$ of the
unitary group $U(p)$ characterized by highest weight components $m_a$
is given by
\beq
C_\ell(R)=C_\ell(\vm)=\sum_{a=1}^p(m_a)^\ell\,\prod_{b\neq a}\left(1-
\frac1{m_a-m_b}\right) \ .
\label{CkUEA}\eeq
Any polynomial in the
Young tableaux row integers can be written as a linear combination
of Casimir invariants (\ref{CkUEA}) of the corresponding representation. The
leading term of the exponential in (\ref{Zpqzkintexpandfinal})
corresponds to the quadratic Casimir $C_2(\vm)$ and yields the expected
commutative result (\ref{Zpqdualcomm}). The higher-order corrections in
$g^2A\,\theta$ involve higher Casimir invariants. Thus the
partition function in the limit of small $g^2A\,\theta$ can be
thought of as coming from the modification of the commutative
Yang-Mills action by an infinite number of higher Casimir
operators and the addition of other non-perturbative
contributions. In commutative lattice gauge theory, higher
Casimir invariants correspond to the presence of higher powers of
the field strength $F$ in the action~\cite{cmr}. The inclusion of
higher Casimir operators in ordinary, two-dimensional Yang-Mills
theory leads to generalized Yang-Mills
theories~\cite{genYM1}--\cite{genYM3} which are defined in the case of
the torus by partition functions of the form
\beq
Z_p\Bigl(A\,,\,\left\{t_\ell\right\}\Bigr)=
\sum_R\exp\left[-\frac A2\,\sum_{\ell>0}t_\ell\,C_\ell(R)\right] \ .
\label{genQCDpart}\eeq
Therefore, the expansion (\ref{Zpqzkintexpandfinal}) presents a
resummation of the long-range effects of noncommutative gauge theory
into local contributions governed by a generalized Yang-Mills theory
(with infinitely-many Casimir operators), plus true non-local
effects which appear non-perturbatively.

\newsection{Polyakov Loop Correlators on the Torus}

We will now proceed to the explicit calculation of observables of the
noncommutative gauge theory. For this, we shall exploit the fact that
spacetime noncommutativity acts in a diagonal way in the instanton
presentation of two-dimensional Yang-Mills theory~\cite{inprep}, i.e.
solutions
of the Yang-Mills equations are modified in a way which preserves
practically all structures present in the commutative theory. The
deformation of the theory due to noncommutativity does not mix
solutions of the equations of motion, nor the components (submodules)
which comprise these solutions. This makes the instanton picture
uniquely well-suited for understanding the noncommutative gauge
theory. The key issue now is to understand how to use this feature to
study Wilson loop correlators of the noncommutative theory.

While exact, group theoretic expressions are available in ordinary $U(p)$
Yang-Mills theory on the torus for the correlators of any number of
Wilson loop observables on arbitrary contours \cite{rusakov},
not all of these are
well-suited for an appropriate transcription to the noncommutative
theory. For instance, topologically trivial Wilson loops lead to
dimension factors in the sums over representations, which hinders a
straightforward Poisson resummation to the instanton
expansion. Furthermore, the fusion numbers for the decomposition of a
direct product into irreducible representations are not known
explicitly for arbitrary $U(p)$ representations. Finally, one needs to
build an expression which is covariant under gauge Morita duality
transformations in order to be able to make the generalization to the
noncommutative case. In this section we will compute a class of
correlation functions in commutative $U(p)$ gauge theory on
${\bf T}^2$ which admits an explicit expansion as a sum over
instanton contributions and which admits a Morita covariant formulation.

\subsection{Physical Correlation Functions\label{PhysCorrFns}}

Consider the Polyakov loop operator \cite{polyakov,suss}
of winding number $m\in\zed$,
\beq
{\sf P}_m(U)=\chi^{~}_{F^{\otimes m}}(U)=\Tr\,U^m \ ,
\label{sfPnU}\eeq
which is defined as the character of the holonomy $U$ of the gauge field
around the inner cycle of the torus in the direct product $F^{\otimes
  m}$, where $F$ is the fundamental representation of the $U(p)$
structure group. From a kinematic point of view, the winding number
$m$ is the length of the single row of boxes in the Young diagram for
the representation $F^{\otimes m}$. The set of all Polyakov loops
(\ref{sfPnU}) in ordinary Yang-Mills theory is thereby a complete set
of characters for the (reducible) representations of the structure
group. They form an algebraic basis which is equivalent to the complete set of
characters of the $\ell^{\,{\rm th}}$ antisymmeterized fundamental
representations $F^{\wedge\ell}$, owing to the relationships \cite{p}
\beq
\chi^{~}_{F^{\wedge\ell}}(U)=\frac{(-1)^\ell}{\ell!}\,\left.
\frac{\partial^\ell}{\partial z^\ell}\exp\left(-\sum_{m=1}^\infty
\frac{{\sf P}_m(U)}m~z^m\right)\right|_{z=0} \ .
\label{chiantisymm}\eeq
All information required about the gauge group and observables
can be written in terms of their correlators, and their winding
numbers give a complete set of labels for representations of the
commutative $U(p)$ gauge group.

{}From a dynamical perspective, the Polyakov loop on a torus
can be thought of as a
physical charge sitting at some point on a spatial circle, with the
integer-valued winding numbers giving the spectrum of colour-electric
charges in the system. Because of the non-vanishing $U(1)$
charge of a single loop, the expectation value of the loop operator
(\ref{sfPnU}) vanishes in the confining $U(p)$ gauge theory, $\langle{\sf
  P}_m(U)\rangle=0$, implying that it would require an infinite amount
of energy to introduce a single fundamental test charge into the
system. By charge conservation, the simplest non-trivial
correlation function involving the operators (\ref{sfPnU}) is the loop
anti-loop correlator~\cite{GPSS}
\beq
W_{p;m}(A_1,A_2)=\Bigl\langle{\sf P}_m(U)\,{\sf P}_{-m}(V)\Bigr\rangle \ .
\label{loopantiloop}\eeq
This correlation function computes the interaction amplitude between a quark
anti-quark pair in the $U(p)$ gauge theory.

To evaluate (\ref{loopantiloop}), we note that the loops split the
torus ${\bf T}^2$ into two cylinders of areas $A_1$ and $A_2$ such that
\beq
A=A_1+A_2
\label{totalarea}\eeq
is the total area of the
torus. By using the cylinder amplitude (\ref{cylinderampl}) we can
thereby write the normalized expectation value (\ref{loopantiloop}) as
\beq
W_{p;m}(A_1,A_2)=\frac1{Z_{p}(A)}\,\int\limits_{U(p)}[\dd U]~[\dd V]~
K_p(A_1;U,V)\,K_p(A_2;V,U)\,{\sf P}_m(U)\,{\sf P}_{-m}(V) \ .
\label{Wncylinder}\eeq
The integrations over the unitary group in (\ref{Wncylinder}) can be
evaluated by noting that products of characters may be written using
multiplicativity and linearity as
\beq
\chi^{~}_R(U)\,\chi^{~}_S(U)=\chi^{~}_{R\otimes S}(U)=
\sum_{R'}N^{R'}_{R,S}~\chi^{~}_{R'}(U) \ ,
\label{charprodN}\eeq
where $N^{R'}_{R,S}$ are the fusion numbers which count the degeneracy
of the irreducible representation $R'$ in the Clebsch-Gordan
decomposition $R\otimes S=\bigoplus_{R'}N^{R'}_{R,S}~R'$. By applying
the orthogonality relations (\ref{charortho}) they may be defined as
\beq
N^{R'}_{R,S}=\int\limits_{U(p)}[\dd U]~\chi^{~}_R(U)\,
\chi^{~}_S(U)\,\chi_{R'}^*(U) \ .
\label{fusiondef}\eeq
By substituting (\ref{cylinderampl}) and (\ref{sfPnU}) into
(\ref{Wncylinder}), and using (\ref{fusiondef}), we arrive at the
group theory presentation
\beq
W_{p;m}(A_1,A_2)=\frac1{Z_{p}(A)}\,\sum_{R,R'}N_{R'\,,\,F^{\otimes m}}^R\,
N^{R'}_{R\,,\,\overline{F}^{\,\otimes m}}~\e^{-\frac{g^2A_1}2\,
C_2(R)-\frac{g^2A_2}2\,C_2(R')} \ .
\label{Wngroupth}\eeq
The fusion numbers for the unitary group are sums of products of
simple delta-functions in the Young tableaux row integers labelling
each irreducible representation. Since the Casimir operators $C_2$ are
quadratic in the row variables, the calculation of the loop
correlators (\ref{Wngroupth}) involves only theta-functions on ${\bf
  T}^2$, to which Jacobi inversion may be applied.

To compute the fusion numbers in (\ref{Wngroupth}), we use
the Weyl formula for the $U(p)$ characters
\beq
\chi^{~}_R(U)=\chi^{~}_{\vm}\left[\e^{2\pi\ii\lambda}\,
\right]=\frac1{\Delta\left[\e^{2
\pi\ii\lambda}\,\right]}~\det_{1\leq a,b\leq p}\,\left(\e^{2\pi\ii m_a
\lambda_b}\right) \ ,
\label{Weylchar}\eeq
where $\e^{2\pi\ii\lambda_a}$, $\lambda_a\in[0,1]$, $a=1,\dots,p$ are
the eigenvalues of the unitary matrix $U$ and
\beq
\Delta\left[\e^{2\pi\ii\lambda}\,\right]=\prod_{a<b}
\left(\e^{2\pi\ii\lambda_a}-\e^{2\pi\ii\lambda_b}\right) \ .
\label{Vandermonde}\eeq
We then transform the unitary integration in
(\ref{fusiondef}) into an integration over the eigenvalues of $U$,
with Jacobian $\Delta[\e^{2\pi\ii\lambda}\,]^2$, and substitute in
(\ref{sfPnU}) and (\ref{Weylchar}) to write (\ref{Wngroupth}) as
\bea
W_{p;m}(A_1,A_2)&=&\frac1{(p!)^2\,Z_{p}(A)}\,\sum_{\vm,\vm^\prime\in\zeds^p}
\e^{-\frac{g^2A_1}2\,C_2(\vm)-\frac{g^2A_2}2\,C_2(\vm')}
\nn\\ && \times\,\prod_{a=1}^p~
\int\limits_0^1\dd\lambda_a~\dd\mu_a~
\left(\,\sum_{c=1}^p \e^{2\pi\ii m\,\lambda_c} \right)
\left(\,\sum_{d=1}^p \e^{-2\pi\ii m\,\mu_d} \right)\nn \\
&&\times\,\det_{1\leq a,b\leq p}\,\left(\e^{2\pi\ii m_a\lambda_b}\right)\,
\det_{1\leq a,b\leq p}\,\left(\e^{-2\pi\ii m_a\mu_b}\right)\nn\\&&\times\,
\det_{1\leq a,b\leq p}\,\left(\e^{2\pi\ii m^\prime_a\mu_b}\right)\,
\det_{1\leq a,b\leq p}\,\left(\e^{-2\pi\ii m^\prime_a\lambda_b}\right) \ .
\label{wraw}\eea
Here we have removed the ordering restriction (\ref{irrepms})
on the integers $m_a$ and $m_a^\prime$ using the permutation symmetry
of (\ref{C2Rn}) and antisymmetry of the determinant factors.
Note that the correlator (\ref{wraw}) is symmetric under the
simultaneous interchange of area labels $A_1\leftrightarrow A_2$ and
reflection of loop winding number $m\to-m$.

The product of determinants in (\ref{wraw}) can be simplified by using the
symmetry of the summand to combine them as
\bea
W_{p;m}(A_1,A_2)&=&\frac{p!}{ Z_{p}(A)}\,\sum_{\vm,\vm^\prime \in\zeds^p}
\e^{-\frac{g^2A_1}2\,C_2(\vm)-\frac{g^2A_2}2\,C_2(\vm')}~
\prod_{a=1}^p~\int\limits_0^1\dd\lambda_a~
\dd\mu_a~\e^{2\pi\ii(m_a - m^\prime_a)\lambda_a }\nn\\&&\times\,
\left(\,\sum_{c=1}^p \e^{2\pi\ii m\,\lambda_c} \right)
\left(\,\sum_{d=1}^p \e^{-2\pi\ii m\,\mu_d} \right)
{}~\det_{1\leq a,b\leq p}\,\left(\e^{
2\pi\ii(m^\prime_a - m_b)\mu_b}\right) \ .
\label{Wnsuminter}\eea
The integrations over $\lambda_a$ in (\ref{Wnsuminter}) can be used to fix the
integers $m_a'$, for each summation index $c$, as $m_a'=m_a+m\,\delta_{a,c}$.
The remaining integrals over $\mu_a$ then produce delta-functions in the
$m_a$,
and thereby reduce (\ref{wraw}) to an expression involving only a single
determinant as
\bea
W_{p;m}(A_1,A_2)&=&\frac{p!}{Z_{p}(A)}~\sum_{\vm \in\zeds^p}~
\sum_{c,d=1}^p\e^{-\frac{g^2A}2\,C_2(\vm)-
\frac{g^2A_1}2\,\bigl[m^2 - 2m\,\bigl(m_c-\frac{p-1}{2}\bigr)\bigr]}\nn\\
&&\times\,\det_{1\leq a,b\leq p}\,\left(\delta_{m_a\,,\,m_b +m\,
\delta_{a,d} -m\,\delta_{a,c}}\right) \ .
\label{Wnsum}\eea
The expression (\ref{Wnsum}) for the loop correlator involves a sum over
$\vm\in\zed^p$ of functions which are symmetric under
permutation of the $m_a$'s along with the determinant of a Kronecker
delta-function. We recall that this feature was also common to the
vacuum amplitude of section~\ref{GroupTheory}, which can be recovered
from the sums in (\ref{Wnsum}) at vanishing loop winding number $m=0$.

Consequently, as with the partition function
in (\ref{Zpcycle}), the correlation function (\ref{Wnsum})
can be rewritten as a sum over conjugacy classes of the symmetric
group $S_p$. With the notion of partition as explained in
section~\ref{GroupTheory}, we can substitute the determinant
expansion (\ref{deltainsert}) into (\ref{Wnsum}) to write it as a sum
over partitions of the form
\bea
W_{p;m}(A_1,A_2)&=&\frac{p!}{Z_{p}(A)}~\sum_{\vm\in\zeds^p}~
\sum_{\stackrel{\scriptstyle{\rm partitions}}{\scriptstyle\vp}}~
\prod_{a=1}^p\frac{(-1)^{\nu_a}}{a^{\nu_a}\,\nu_a!}~
\prod_{k=1}^{|\vnu|}\e^{-\frac{g^2A}2\,p_k\bigl(m_k-\frac{p-1}{2}
\bigr)^2}\nn\\&&\times\,\sum_{l=1}^{|\vnu|}p_l~\sum_{b=1}^{p_l}
\e^{-\frac{g^2A_1}2\,b\,\bigl[m^2-2m\,\bigl(m_l-\frac{p-1}2
\bigr)\bigr]-\frac{g^2A_2}2
\, (b-1) \bigl[m^2-2m\,\bigl(m_l-\frac{p-1}2\bigr)\bigr]}  \ . \nn\\&&
\label{W1pcycle}\eea
Here and in the following we assume that $m\neq0$. We have
expressed the sums over the indices $c$ and $d$ in (\ref{Wnsum}) as a
sum over components $b$ of distinct cycles in a partition. As before,
the factors of $p_k$ arise from the grouping of terms into cycles
which subtracts off contributions whose row variables $m_k$ coincide.

To extract physical information from the correlator (\ref{W1pcycle}), we set
$m=1$ and take the size of the torus to infinity while holding fixed one
of the areas, say $A_1$~\cite{GPSS}. This damps out all flux in the
area $A_2$ leaving only
colour-electric flux in $A_1$, and the correlator gives the binding
energy of a quark anti-quark pair interacting on the plane
$\real^2$. In this limit, only the lowest (singlet) representation
survives the sums in (\ref{Wnsum}), and one finds
\beq
\lim_{A_2\to\infty}\,W_{p;1}(A_1,A_2)=p^2~\e^{-g^2 A_1 p(p^2 +11)/12} \ .
\label{Wp1largeA2}\eeq
Restoring a contribution to the quadratic Casimir which we have
neglected,
we can identify a linear confining potential between the
fundamental representation quark anti-quark
pair with familiar string tension $g^2 p$.

\subsection{Instanton Contributions}

To rewrite (\ref{W1pcycle}) as a sum over instantons, we carry out
Poisson resummations of the integers $m_k$ using (\ref{Poisson}). The
result for the sum over the dual integers $q_k$, restricted to
principal $U(p)$ bundles over ${\bf T}^2$ of fixed total Chern number
$q=\sum_kq_k$, allows one to define the correlation functions of
Polyakov loop operators in Yang-Mills theory on a projective
module of fixed topological numbers $(p,q)$. By incorporating the
constraint on $\sum_kq_k$ in the definition of partition in
(\ref{partitiondef}) for $\theta=0$, we can thereby write
\bea
W_{p,q;m}(A_1, A_2) &=& \frac{p!}{Z_{p,q}(A,0)}~
\sum_{\stackrel{\scriptstyle{\rm partitions}}{\scriptstyle (\vp,\vq)} }
{}~\frac{(-1)^{|\vnu|}}{\prod_a\nu_a!}~\prod_{k=1}^{|\vnu|}\,
\sqrt{\frac{2\pi^2}{g^2A\,(p_k)^3}}~
\e^{-\frac{2 \pi^2}{g^2 A}\,\frac{(q_k)^2}{p_k}}\nn\\
&&\times\,\sum_{l=1}^{|\vnu|}p_l~\sum_{b=1}^{p_l}
\e^{\frac1{2 A p_l}\,\bigl(bA_1 + (b-1) A_2\bigr)\bigl[g^2m^2\bigl(
bA_1 +(b-1)A_2  - p_l A\bigr) - 4 \pi \ii m\,q_l\bigr]} \ . \nn\\&&
\label{Wpcycle}\eea
The sum over partition components in (\ref{Wpcycle}) gives the classical
contribution to the Polyakov loop, so that (\ref{Wpcycle}) can be
interpreted as the statistical average over all instantons of the
Polyakov loop. Note that for the $U(p)$ gauge theory on the torus
defined in a sector of fixed, non-trivial 't~Hooft flux $q$,
Polyakov loop operators (\ref{sfPnU}) should be
accompanied by the transition functions $(\Gamma_q)^m$,
\beq
{\sf P}_{q;m}(U)=\Tr\,U^m\,\left(\Gamma_q\right)^m \ .
\label{Polyakovtwist}\eeq
Here $\Gamma_q$ is the corresponding $SU(p)$ twist-eating matrix along the
inner cycle of ${\bf T}^2$ and its inclusion is necessary in order
to ensure invariance of the correlation functions under large gauge
transformations~\cite{sz1,AMNS,gt}.

In order to make contact with the noncommutative theory later on,
we will require objects which transform covariantly under Morita equivalence.
Such an object can be constructed from the correlator (\ref{Wpcycle}) if we
carry out a Fourier transform in the direction transverse to the loop
windings.
For this, on a torus of fixed total area (\ref{totalarea}),
we set $A_1 = (1-x)A$, $A_2 =xA$, and integrate over $x\in[0,1]$. The result
of this transform yields the pair correlator of Polyakov loops with winding
number $m$ and momentum $n\in\zed$ as
\beq
W_{p,q;m,n}(A,0)=\int\limits_0^{1}\dd x~\e^{-2 \pi \ii n\,x}~
W_{p,q;m}\Bigl( (1-x) A\,,\, x A\Bigr) \ .
\label{Fourierloop}\eeq
Upon substitution of (\ref{Wpcycle}) into (\ref{Fourierloop}), the integral
over $x$ can be evaluated in terms of error functions with the
result
\bea
W_{p,q;m,n}(A,0)&=&\frac{p!}{Z_{p,q}(A,0)}~
\sum_{\stackrel{\scriptstyle{\rm partitions}}{\scriptstyle (\vp,\vq)} }
\frac{(-1)^{|\vnu|}}{\prod_a\nu_a!}~\prod_{k=1}^{|\vnu|}\,
\sqrt{\frac{2\pi^2}{g^2A\,(p_k)^3}}~
\e^{-\frac{2 \pi^2}{g^2 A}\,\frac{(q_k)^2}{p_k}}\nn\\
&&\times\, \sum_{l=1}^{|\vnu|}
\,\left(p_l\right)^2~\sqrt{\frac{2\pi^2}{g^2 A\,m^2 p_l}}~
\exp\left[\frac{\Bigl(4\pi(m\,q_l-n\,p_l) - \ii g^2 A\,m^2 p_l
\Bigr)^2}{8 g^2 A\,m^2 p_l}\,\right]\nn \\
&& \times\,\sum_{b=1}^{p_l}\ii\left({\rm erf}
\left[\frac{ 4 \pi\,(m\,q_l-n\,p_l) - \ii g^2 A\,m^2 (2 - 2b+p_l) }
{\sqrt{8 g^2A\,m^2 p_l} } \right] \right. \nn \\&&-\left.
{\rm erf}\left[\frac{ 4 \pi\,(m\,q_l-n\,p_l) - \ii g^2 A\,m^2(p_l-2b) }
{\sqrt{8 g^2 A\,m^2 p_l}} \right] \right) \ .
\label{WpqwnA0}\eea
Note that the sum in (\ref{WpqwnA0}) over components $b$ of each cycle
in a given partition is telescopic and reduces to the initial and
final terms in the series.

By discarding irrelevant constants, the final expression for the Polyakov
loop correlator thereby reads
\bea
W_{p,q;m,n}(A,0)&=&\frac1{Z_{p,q}(A,0)}~
\sum_{\stackrel{\scriptstyle{\rm partitions}}{\scriptstyle (\vp,\vq)} }
\frac{(-1)^{|\vnu|}}{\prod_a\nu_a!}~\prod_{k=1}^{|\vnu|}\,
\sqrt{\frac{2\pi^2}{g^2A\,(p_k)^3}}~
\e^{-\frac{2 \pi^2}{g^2 A}\,\frac{(q_k)^2}{p_k}}\nn\\
&&\times\, \sum_{l=1}^{|\vnu|}
\,\left(p_l\right)^2~\sqrt{\frac{2\pi^2}{g^2 A\,m^2 p_l}}~
\exp\left[\frac{\Bigl(4\pi\,
{m\choose n}\wedge{p_l \choose q_l}-\ii g^2 A\,m^2 p_l
\Bigr)^2}{8 g^2 A\,m^2 p_l}\,\right]\nn \\&&\times\,{\rm Im}\left(
{\rm erf}\left[\frac{4 \pi\,
{m\choose n}\wedge{p_l \choose q_l}+ \ii g^2A\,m^2p_l}
{\sqrt{8 g^2A\,m^2 p_l} } \right]  \right)  \ ,
\label{resummedW}\eea
where we have defined the cross-product of two-dimensional integer
vectors by the determinant
\beq
{a\choose b}\wedge{a'\,\choose b'}=\det\left(\,\begin{matrix}a&a'\\b&b'
\end{matrix}\,\right) \ .
\label{2Dcrossprod}\eeq
The symmetry of the correlator under interchange of areas $A_1 \leftrightarrow
A_2$ and $m\to-m$ is manifest in the momentum form (\ref{resummedW}), whereby
the exchange of area labels is replaced by a reflection of momentum
$n\to-n$. Note that, like the restriction to sectors of fixed magnetic
flux $q$, the transverse Fourier transform of the loop correlator is
superfluous here, because the physical correlation functions can always be
recovered as the Fourier series
\beq
W_{p;m}(A-A_2,A_2)=\sum_{q=-\infty}^\infty\e^{\pi\ii(p-1)\,q}~
\sum_{n=-\infty}^\infty\e^{2\pi\ii n\,A_2/A}~W_{p,q;m,n}(A,0) \ .
\label{physcorrrecover}\eeq
In other words, the translation group of the torus, with $U(1)$
characters labelled by the transverse momenta $n$, is completely decoupled
from the structure group, with characters labelled by the
winding numbers $m$. In this case the translation and rotation groups appear
together only in a free, direct product decomposition.

The precise meaning of the correlation functions (\ref{resummedW}) is
most naturally understood in the noncommutative gauge theory. In that
case, the separation of colour and translational degrees of freedom is
not possible, reflecting the inherent non-locality of the quantum
field theory. Similarly to the case of the vacuum energy, the pair
correlator in the noncommutative case only admits an unambiguous
definition on a fixed projective module and for fixed momentum
$n$. The physical meaning of this restriction will be elucidated in
the next section.

\newsection{Open Wilson Line Correlators\label{OWLC}}

In this section we shall transcribe the results obtained in the
previous subsection into the noncommutative gauge theory by adapting
the technique of~\cite{inprep} to correlation functions. While the physical
meaning of Polyakov loop correlators was discussed in
section~\ref{PhysCorrFns}, their noncommutative cousins are a bit more
intricate to describe. We will
therefore begin with a description of what sort of physics will be
described by the noncommutative correlation functions, and then present the
exact analytic expressions for the pair correlators in this case.

\subsection{Kinematics of Noncommutative Dipoles}

The simplest way to understand the necessity for the transverse
momentum label of correlators is to look at the Dirac quantization condition
for electric flux in the noncommutative field
theory~\cite{HofVer,ks1}. Let us briefly recall how this works, in a
manner tailored to what we shall need in the following. The winding
number $m\in\zed$ of a Polyakov loop (\ref{sfPnU}) corresponds to the
large gauge transformation $A\mapsto A+2\pi\,m$ around the inner cycle
of the torus ${\bf T}^2$. These transformations are generated in field
space by the electric field operator $E=\delta/\delta A$ and can be
realized in the commutative case by the $U(1)$ gauge transformation
\beq
\Omega(y)=\e^{2\pi\ii y} \ ,
\label{Egaugetransf}\eeq
with $y\in[0,1]$ the coordinate along the inner cycle.

However, in the noncommutative case the corresponding star-gauge
transformation
generated by (\ref{Egaugetransf}) also acts on the non-zero modes of the gauge
fields and would generate a translation in the spatial dependence $x\mapsto
x-\theta\,m$ of the gauge field, where the coordinate $x\in[0,1]$ runs
transverse to $y$. Requiring that states in the physical Hilbert space of the
noncommutative field theory be gauge invariant thereby
requires also a simultaneous overall translation in space, which may be
implemented by the transverse momentum operator $P$. This is guaranteed if the
zero modes of the operator $E+\theta\,P$, rather than the
electric field itself, are integral. Therefore, the modified Dirac
quantization condition on the electric field is
\beq
\int\limits_0^1\dd x~\dd y~\Tr\,E=m-n\,\theta\equiv e_{m,n} \ ,
\label{Diracquanttheta}\eeq
where $m\in\zed$ and
\beq
n=\int\limits_0^1\dd x~\dd y~\Tr\,P~\in~\zed
\label{momzeromode}\eeq
is the zero-mode of the transverse momentum operator. Note that since $P$
generates geometrical translations of the torus, its zero-mode spectrum is
still integral. The change (\ref{Diracquanttheta}) in the spectrum of
colour-electric charges $m$ is completely analogous to the way in which the
commutative ranks $p$
are modified to non-integer module dimensions $p-q\,\theta$ in the
noncommutative case. It shows explicitly how Young tableaux row
integers are modified to non-integer ones determined by the transverse
translational zero-modes. Note that the gauge covariant momentum can be
represented by the operator $F\,E$, whose zero mode spectrum is given by
$n\,(p-q\,\theta)$.

The modified spectrum (\ref{Diracquanttheta}) of electric charges
is due to the fact that the elementary quanta of the
noncommutative gauge theory are described by weakly-interacting,
non-local electric dipoles~\cite{Sheikh1,BigSuss}, whose dipole
moments $e_{m,n}$ are related to their center-of-mass momenta $n$
through the relation (\ref{Diracquanttheta}). The dipole's motion
is transverse to its extension as a rigid rod along the
$y$-direction of ${\bf T}^2$. When $\theta=0$ they become ordinary
point-like quanta, with a complete decoupling of momentum and
winding number. Similarly to the way that the Polyakov loops
(\ref{sfPnU}) correspond to fundamental charges on the torus, the
dipole excitations are created and annihilated by the open Wilson
line operators~\cite{IIKK} \footnote{ Other aspects of open Wilson
line operators may be found in \cite{AMNS,rey1,rey2,ghi}.} \beq
{\sf O}_{m,n}[U]=\int\limits_0^1\dd x~\dd y~
\Tr\,U_\star(x,y\,;\,x,y+m-n\,\theta)\star\e^{-2\pi\ii n\,x} \ ,
\label{OpenWilsondef}\eeq where $U_\star(x,y\,;\,x,y+e_{m,n})$ is
the noncommutative parallel transport operator defined along the
straight open contour starting at the point $(x,y)\in{\bf T}^2$,
winding $m$ times around the $y$-direction, and then ending up at
the shift $-n\,\theta$ from $y$. They implicitly include the
path-ordered star-exponentials of the appropriate background
Abelian gauge field required to absorb the large gauge
transformation which is present due to the non-vanishing magnetic
charge $q$. The operators (\ref{OpenWilsondef}) are
gauge-invariant but non-local. They also illustrate, via the
requirement of star-gauge invariance, the necessity of the
transverse Fourier transform in the noncommutative setting.

The energy of an electric dipole of moment
(\ref{Diracquanttheta}) in the gauge background characterized by the
topological numbers $(p,q)\in\zed^2$ can be obtained from the
perturbative spectrum of the quantum Hamiltonian on $\real\times{\bf
  T}^2$. At leading order in perturbation theory, one finds the
ground state energy~\cite{ks2}
\beq
{\cal E}_{p,q;m,n}(A,\theta)=\frac1{p-q\,\theta}\,\left[
\frac{g^2A}2\,\Bigl(m-n\,\theta\Bigr)^2+\left|{m\choose n}
\wedge{p\choose q}\right|\,\right] \ .
\label{BPS14energy}\eeq
With our choice of subtraction of background flux in (\ref{Zpqfnint}),
the instanton contribution to this energy from the constant curvature of the
Heisenberg module itself vanishes as in (\ref{Zpqpartitionsum}), because
this sets the energy of the stable vacuum state (without dipole excitations)
to zero, ${\cal E}_{p,q;0,0}(A,\theta)=0$. The first term in
(\ref{BPS14energy}) is the energy associated to zero modes and
is simply the kinetic energy corresponding to the dipole moment
$e_{m,n}$. The second term is the contribution from oscillatory modes and
it gives the energy of the massless excitations of the
dipole. It coincides with the zero mode spectrum of the operator $\Tr FE-\Tr
F\,\Tr E$ on ${\bf T}^2$.

In the case of supersymmetric Yang-Mills theory on the noncommutative
torus, the energy formula (\ref{BPS14energy}) can be computed directly
from the central charges of the corresponding BPS algebra~\cite{ks1}. It is an
exact result in this case and corresponds generically to a $\frac14$-BPS
state. These states also naturally arise in string theory, wherein the integer
pair $(p,q)$ represents (D0,D2) brane charges on ${\bf T}^2$ in gauge
backgrounds, while $(m,n)$ corresponds to the winding numbers and
momenta of fundamental strings wrapping around the cycles of the
torus. For $\frac14$-BPS states, the string charges $K$
must decompose into the particle charges as $K=mq-np={m\choose
  n}\wedge{p\choose q}$~\cite{ObersPio}.

\subsection{Exact Dipole Interaction Amplitudes}

We will now turn to the dynamics of electric dipoles in the
noncommutative gauge theory. The interaction energy between a pair of dipoles
can be computed by using the open Wilson lines (\ref{OpenWilsondef}),
along with the conservation of momentum and electric charge in the gauge
theory to define
\beq
\Bigl\langle{\sf O}_{m,n}[U]\,{\sf O}_{m',n'}[V]\Bigr\rangle\equiv
\delta_{m,m'}\,\delta_{n,n'}~W_{p,q;m,n}(A,\theta) \ .
\label{OpenWilsonintdef}\eeq
When $\theta=r/s$ is a rational number, Morita duality provides a one-to-one
correspondence between the Polyakov loop operators (\ref{Polyakovtwist}) in
commutative Yang-Mills theory and the noncommutative open Wilson line
operators (\ref{OpenWilsondef})~\cite{sz1,AMNS,gt}. We can use this
covariance to explicitly deduce the noncommutative version
(\ref{OpenWilsonintdef}) of the amplitude (\ref{resummedW}). Under the
gauge Morita equivalence transformations parameterized by the
$SL(2,\zed)$ matrices (\ref{SL2Zcommtransf}), the pairs of integers
$(p_k,q_k)$ transform as $SL(2,\zed)$ doublets (\ref{pqSL2Zdoublets}),
while the dimensionless Yang-Mills coupling constant $g^2A$ transforms
according to (\ref{dimcouplingtransf}). Looking at (\ref{resummedW}),
we see that the combinations of variables $g^2 A\,m^2 p_k$ and ${m\choose
  n}\wedge{p_k\choose q_k}$ play a special role. Requiring these two
sets of combinations to be Morita invariant is tantamount to demanding that
the integer $m$ transform like the dimension of a module and that the
cross-product ${m\choose n}\wedge{p_k\choose q_k}$ be invariant.
These requirements are met if the pair of integers $(m,n)$ has the expected
vector transformation rule under $SL(2, \zed)$ as in (\ref{nmSL2Zdoublets}),
\beq
\begin{pmatrix}\,m'\,\cr\,n'\,\cr\end{pmatrix}=
\begin{pmatrix}\,s&-r\,\cr\,r'&s'\,\cr\end{pmatrix}
\begin{pmatrix}\,m\,\cr\,n \,\cr\end{pmatrix} \ ,
\label{mnSL2Zdoublet}\eeq
since the cross-product (\ref{2Dcrossprod}) is clearly an $SL(2,\zed)$
invariant.

With the identification of the transformation properties of (\ref{resummedW})
under gauge Morita equivalence, we can now apply the procedure that
was used in~\cite{inprep} to determine the partition function of
two-dimensional Yang-Mills theory on a noncommutative torus. The
Morita transformations (\ref{pqSL2Zdoublets}) and
(\ref{mnSL2Zdoublet}) immediately provide the result for all tori with
rational noncommutativity parameters $\theta =r/s$ through the
replacement of dimensions $p_k$ and electric flux $m$ with their
noncommutative generalizations $p_k-q_k\,\theta$ and
$m-n\,\theta$. The definition of a partition $(\vp,\vq)$ is modified
to its full noncommutative form (\ref{partitiondef}), and the
components of a partition are identified with the topological numbers
of projective modules. The Morita transform of the pair correlator
(\ref{resummedW}) thereby gives
\bea
W_{p,q;m,n}(A,\theta)&=&\frac{1}{Z_{p,q}(A,\theta)}~
\sum_{\stackrel{\scriptstyle{\rm partitions}}
{\scriptstyle(\vp,\vq)}}\,\frac{(-1)^{|\vnu|}}{\prod_a\nu_a!}\,
\prod_{k=1}^{|\vnu|}\sqrt{\frac{2\pi^2}{g^2A\,
(p_k-q_k\,\theta)^3}}\non&&\times\,\exp\left[-\frac{2\pi^2}{g^2A}\,
\sum_{k=1}^{|\vnu|}\left(p_k-q_k\,\theta\right)
\left(\frac{q_k}{p_k-q_k\,\theta}-\frac q{p-q\,\theta}\right)^2\right]
\nn  \\&& \times\,\sum_{l=1}^{|\vnu|}
\left(p_l- q_l\,\theta\right)^2~\sqrt{\frac{2 \pi^2}{g^2 A\,(m-n\,\theta)^2
\left(p_l - q_l\,\theta\right)}}\nn\\&&\times\,
\exp\left[\frac{\Bigl(4 \pi\,{m\choose n}\wedge{p_l\choose q_l}-
\ii g^2 A\,(m-n\,\theta)^2\left(p_l -q_l\,\theta
\right)\Bigr)^2}{8 g^2 A\,(m-n\,\theta)^2\left(p_l -q_l\,\theta\right)
}\,\right]\nn \\&& \times\,{\rm Im} \left(
{\rm erf}\left[\frac{ 4 \pi\,{m\choose n}\wedge{p_l\choose q_l}+
\ii g^2 A\,(m-n\,\theta)^2\left(p_l -q_l\,\theta\right)}
{\sqrt{8 g^2A\,(m-n\,\theta)^2
\left(p_l -q_l\,\theta\right)} } \right]  \right) \ ,
\label{wtheta}\eea
where we have taken into account the shift of the instanton
contributions to the action in (\ref{resummedW}) by the constant
curvature (\ref{bpq}) associated with the topological numbers
$(p,q)\in\zed^2$,
which is produced by the Morita mapping. With the vector $SL(2,\zed)$
transformation properties of $(\vp,\vq)$ and $(m,n)$, the correlator
(\ref{wtheta}) transforms as the square of a module dimension
$(p-q\,\theta)^2$, as it should since it involves a double trace.

As was noted in~\cite{inprep} for the case of the vacuum energy, the
expression (\ref{wtheta}) for the correlation function at rational
values of $\theta$ can be naturally generalized to irrational $\theta$.
Since the correlation function can be expressed as a sum over contributions
from solutions of the classical equations of motion parameterized by
partitions $(\vp,\vq)$, the arguments given in~\cite{inprep} for the
continuity in $\theta$ of the localization of the path integral onto classical
gauge field configurations apply here as well. These arguments strongly
suggest that (\ref{wtheta}) defines the pair correlation functions of open
Wilson line operators for {\it all} values of $\theta$. In the
following we will assume this to be the case. With this we require that
$m-n\,\theta\neq0$. In what follows it will prove more convenient to
write the general correlator (\ref{wtheta}) in terms of an integral
representation for the error function as
\bea
&&W_{p,q;m,n}(A,\theta)~=~\frac{1}{Z_{p,q}(A,\theta)}~
\sum_{\stackrel{\scriptstyle{\rm partitions}}
{\scriptstyle(\vp,\vq)}}\,\frac{(-1)^{|\vnu|}}{\prod_a\nu_a!}\,
\prod_{k=1}^{|\vnu|}\sqrt{\frac{2\pi^2}{g^2A\,
(p_k-q_k\,\theta)^3}}\non&&~~~~~~~~~~\times\,\exp\left[-\frac{2\pi^2}{g^2A}\,
\sum_{k=1}^{|\vnu|}\left(p_k-q_k\,\theta\right)
\left(\frac{q_k}{p_k-q_k\,\theta}-\frac q{p-q\,\theta}\right)^2\right]
\nn  \\&&~~~~~~~~~~\times\,\sum_{l=1}^{|\vnu|}
\frac{\left(p_l-q_l\,\theta\right)^2}2\,
\int\limits_{-1}^1\dd s~ \e^{ \pi \ii(1-s)\,{m\choose n}\wedge
{p_l\choose q_l}-g^2 A\,(m-n\,\theta)^2
(p_l - q_l\,\theta)(1-s^2)/8} \ .
\label{wthetaint}\eea

\subsection{Dipole Vacua}

Armed with the exact analytic expressions for the open Wilson lines,
in the subsequent sections we shall describe some of their physical
properties. The pair correlators (\ref{wtheta}) and (\ref{wthetaint})
naturally
involve, for each component $(p_k,q_k)$ of each partition $(\vp,\vq)$,
both the associated dipole kinetic energy and massless oscillations that
appear in the energy formula (\ref{BPS14energy}). As we will see, the latter
contributions play a prominent role in the instanton expansion, and so
we will first briefly discuss in this subsection some further aspects
of them. What will be particularly important in the following is that
in the dipole vacuum state whereby the oscillatory modes vanish,
\beq
{m\choose n}\wedge{p\choose q}=0 \ ,
\label{dipolevaccond}\eeq
the dipole kinetic energy corresponds to a maximally supersymmetric
$\frac12$-BPS state~\cite{ks1}, just like the instanton configuration
$(\vp,\vq)=(p,q)$ corresponding to the constant curvature gauge fields
of the projective module. From (\ref{dipolevaccond}) it follows that
the transverse dipole momentum of this state is given by $n=n_0$, where
\beq
n_0=e_{m,n_0}~b_{p,q} \ .
\label{dipolemomEM}\eeq
The right-hand side of (\ref{dipolemomEM}) is just the Poynting vector
of the gauge background with constant magnetic flux (\ref{bpq}) and
noncommutative electric flux (\ref{Diracquanttheta}). In other words,
the dipole vacuum is a maximally supersymmetric state characterized by
the condition that its transverse momentum cancels exactly the
momentum of the background electromagnetic field in the gauge vacuum.

As the cross-products in (\ref{wtheta}) and (\ref{wthetaint}) produce
gauge Morita invariants, it is natural to identify the product as a
$\zed$-bilinear pairing between the Abelian group of
topological numbers $(p_k,q_k)$ and the dual group consisting of
geometrical loop winding and momentum numbers $(m,n)$. In other words,
we regard the product as a $(1,1)$ form on K-theory
\beq
\wedge\,:\,{\rm K}_0^*\times{\rm K}^{~}_0~
\longrightarrow~\zed \ ,
\label{crosspairing}\eeq
where ${\rm K}_0^*$ is the Fourier dual of the ${\rm K}_0$ lattice
obtained via Poisson resummation as in section~\ref{InstExp}. Later on
we shall see more evidence to support this claim. It is tempting to
identify ${\rm K}_0^*={\rm K}^0$ and (\ref{crosspairing}) as the
natural pairing between K-homology and K-theory. This would identify
the pairs of integers $(m,n)$ as the topological numbers corresponding
to homotopy classes of Fredholm operators and the cross-product
as the index map. The vanishing condition (\ref{dipolevaccond}) would
then imply the matching of chiral and anti-chiral zero modes of an
associated Fredholm operator, as we expect for a state of maximal
supersymmetry. Since this description is valid for any $\theta$, it
explicitly demonstrates, through Morita duality, why the
cross-products play a role in the commutative correlation functions
(\ref{resummedW}), even though their momentum dependence is
superfluous. However, within the context of two-dimensional
noncommutative gauge theory, the dipole picture will prove to be the
more descriptive one. Indeed, as we will see, the pairs $(m,n)$ are
most naturally understood within a group theoretic setting analogous
to those of section~\ref{InstExp}, in which the cross-products won't
explicitly appear.

\newsection{Moduli Dependence of Dipole Correlation Functions}

In this section we will analyse the behaviours of the correlators
obtained in the previous section in various limits of the coupling
parameters of the noncommutative gauge theory, and thereby extracting their
physical significance. In particular, we will develop the deformed
group theory representation of the correlators and use this to justify
the interpretations we have made previously, most notably in
section~\ref{InstExp}. The connections with the commutative
calculations of Polyakov loop correlators will thereby illuminate the group
theoretic interpretation of two-dimensional noncommutative Yang-Mills
theory.

\subsection{Weak-Coupling Limit}

The weak-coupling limit $g^2 A \rightarrow 0$
of two-dimensional Yang-Mills theory defines a topological field theory
\cite{witten92}.
This limit provides a method to not only calculate correlation functions
in the topological field theory, but also to identify fundamental structures
in the physical gauge theory. For example, in~\cite{inprep} the
weak-coupling limit of the partition function was used to identify the
moduli space of instantons for gauge theory on the noncommutative torus.
In this subsection we will point out some properties of the weak-coupling
limit of the open Wilson line correlator (\ref{wtheta}).
This limit is most easily determined by using the
integral representation (\ref{wthetaint}).

As in the case of the vacuum amplitude (\ref{Zpqpartitionsum}), in the
weak-coupling limit the sum over partitions in (\ref{wthetaint}) is
exponentially dominated by the partition containing only a single
submodule, namely the projective module on which the noncommutative
Yang-Mills theory is defined, $(\vp,\vq)=(p,q)$. Consequently, the
leading behaviour of the correlator (\ref{wthetaint}) at weak-coupling
is given by
\bea
\lim_{g^2 A \rightarrow 0}\,W_{p,q;m,n}(A,\theta)&=&
\frac{(p-q\,\theta)^2}2\,
\int\limits_{-1}^1\dd s~ \e^{\pi\ii(1-s)\,{m\choose n}\wedge{p
\choose q}-g^2 A\,(m-n\,\theta)^2
(p -q\,\theta)(1-s^2)/8}\non&&+\,O\left( \e^{-1/g^2 A}\right) \ .
\label{weakcouplingcorr}\eea
The integration in (\ref{weakcouplingcorr}) depends crucially on
whether or not the cross-product vanishes, as in
(\ref{dipolevaccond}), and so we must consider these two distinct cases
separately.

First of all, a power series expansion in $g^2 A$
for non-vanishing ${m\choose n}\wedge{p\choose q}$ gives
\bea
\lim_{g^2 A \rightarrow 0}\,W_{p,q;m,n}(A,\theta)&=&
\frac{ g^2 A\,(m-n\,\theta)^2
(p-q\,\theta)^3 }{4 \pi^2\,\left[{m\choose n}\wedge{p\choose q}
\right]^2}+O\Bigl(\left(\,g^2A\right)^2\Bigr)
\non&&+\,O\left(\e^{-1/g^2 A}\right)
{}~~~~{\rm for}~~{m\choose n}\wedge{p\choose q}\neq0 \ ,
\label{wcnv}\eeq
and so in this case the correlator vanishes linearly with the
dimensionless coupling constant. On the other hand, if
(\ref{dipolevaccond}) holds, then the correlator has a non-vanishing
weak-coupling limit given by
\bea
\lim_{g^2 A \rightarrow 0}\,W_{p,q;m,n}(A,\theta)&=&
(p-q\,\theta)^2\,\left[1 - \frac{g^2 A}{12}\,\Bigl(m-n\,\theta
\Bigr)^2\Bigl(p  -q\,\theta\Bigr)+O\Bigl(\left(\,g^2A\right)^2\Bigr)
\right]\non&&+\,O\left( \e^{-1/g^2 A}\right)~~~~{\rm for}~~
{m\choose n}\wedge{p\choose q}=0 \ .
\label{wcv}\eea
{}From these two cases we see the significance of the dipole vacuum
state condition (\ref{dipolevaccond}). The most intriguing aspect of
these results is that they are essentially independent
of the noncommutativity parameter $\theta$. In particular,
we see that even in the commutative theory there is a fundamental
importance to the cross-product ${m\choose n}\wedge{p\choose q}$ which
brings together loop winding number $m$ and momentum $n$.

This analysis leads to a physical interpretation of the
cross-product in the commutative theory, i.e. without recourse to
dipole physics. The weak coupling-limit projects the quantum gauge theory onto
a particular topological sector characterized by K-theory charges
$(p,q)\in\zed^2$. This sector can support certain colour-electric
charges and loop momenta $(m,n)\in\zed^2$. By ``support'' we mean
that the projected theory can form a gauge-invariant singlet given the
specific loop configuration. From (\ref{wcnv}) and (\ref{wcv}) we see
that ${m\choose n}\wedge{p\choose q}$ is exactly the parameter
which determines if a singlet configuration is available for loops
$(m,n)$. If not, then the condition (\ref{dipolevaccond}) is violated
and the pair correlator vanishes. It is amusing to note that the
weak-coupling limit of a loop configuration $(m,n)$ is identical to
that of $(k\,m\,,\,k\,n)$ for any integer $k\neq 0$. In this way the
correlation function fixes a topological equivalence relation
$(m,n) \sim (k\,m\,,\,k\,n)~~\forall k\in\zed-\{0\}$. This equivalence comes
about from the fact that the singlet condition requires the integer
vectors ${m\choose n}$ and ${p\choose q}$ to be parallel in the
cross-product (\ref{2Dcrossprod}), so that the loops are
commensurate with the bundle over the torus on which the gauge theory
is defined.

In the physical commutative gauge theory, the group theory expansion
(\ref{Wngroupth}) shows explicitly that the pair correlator
always returns the result (\ref{wcv}) at $\theta=0$ in the weak-coupling
limit.
This is because the physical gauge theory is defined as the double
sum (\ref{physcorrrecover})
over Chern number $q$ and loop momentum $n$.  There are
always terms in these series with $(p,q)$ and  $(m,n)$ satisfying the
condition (\ref{dipolevaccond}).
Nevertheless, this property clearly demonstrates the importance of
the loop momentum $n$, and the  cross-product ${m\choose
  n}\wedge{p\choose q}$, in the restriction of the commutative gauge
theory to a sector of fixed magnetic charge $q$. The theory defined on
a fixed projective module $(p,q)$, even in the commutative case, has a natural
set of observables which are necessarily labelled by {\it two}
integers $(m,n)$, rather than just one.

Going back to the dipole interpretation, we recall that the condition
(\ref{dipolevaccond}) turned the $\frac14$-BPS dipole state into a
$\frac12$-BPS state. This is very natural given that the constant
curvature condition defines $\frac12$-BPS gauge field configurations.
The weak-coupling limit in this way projects the
full quantum field theory onto the Higgs branch of the moduli space of
$\frac12$-BPS solutions. The entire moduli space in the corresponding
supersymmetric gauge theory may be viewed as a fibration over this
branch. Maximal supersymmetry in this way controls which dipole
configurations have non-trivial topological interactions. Alternatively, we
may
regard (\ref{dipolevaccond}) as a requirement of momentum
conservation. Only when the total momentum of the dipole and
background electromagnetic field configuration is conserved do we obtain a
non-vanishing scattering amplitude.

\subsection{Strong-Coupling Expansion}

The integral representation of the correlation function
(\ref{wthetaint}) allows us to carry out a resummation procedure
analogous to that performed on the partition function in
section~\ref{DualExp}. The result is a form of the correlation function
which is amenable to investigations of the strong-coupling limit and
physical interpretation of the correlators in noncommutative
Yang-Mills theory. The calculation is identical to that of the
partition function and so we shall only sketch the derivation.

We begin from (\ref{wthetaint}) by imposing the constraints
(\ref{partitiondef}) defining partitions directly as in
(\ref{Zpqunconstr}), and then again Fourier resolve the
delta-functions over the topological charges $q_k$ and
module dimensions $z_k$ as in (\ref{Zpqdeltaresolve}) to get
\bea
W_{p,q;m,n}(A,\theta) &=& \frac{1}{Z_{p,q}(A,\theta)}~
\sum_{\vp,\vq}\frac{(-1)^{|\vnu|}}{\prod_a\nu_a!}~
\delta^{~}_{p\,,\,\sum_kp_k}~\int\limits_0^1 \dd \lambda ~
\e^{- 2 \pi \ii \lambda\,q}\nn\\&&\times\,\prod_{k=1}^{|\vnu|}\,
\sqrt{\frac{2\pi^2}{g^2A}}~\int\limits_{0^+}^\infty\frac{\dd z_k}
{(z_k)^{3/2}}~\int\limits_{-\infty}^{\infty} \frac{\dd x_k}{2\pi}~
\e^{\ii x_k (z_k-p_k+q_k\,\theta)}~\e^{2 \pi \ii \lambda
\,q_k-\frac{2\pi^2}{g^2A}\,\frac{(q_k)^2}{z_k}}\nn  \\
&&\times\,\sum_{l=1}^{|\vnu|}\frac{(z_l)^2}2~
\int\limits_{-1}^1\dd s~ \e^{ \pi
  \ii(1-s)\,\bigl(q_l(m-n\,\theta)-n\,z_l\bigr) -
g^2 A\,z_k (m-n\,\theta)^2(1-s^2)/8} \ .
\label{correlresolve}\eea
The $q_k$ series can be Poisson resummed and the resulting Gaussian
integrations over the $x_k$ variables carried out. The results are identical
to what was found for the partition function in (\ref{Zpqqtom}) except
for a factor containing information about the Wilson lines. After a
subsequent Poisson resummation of the $p_k$ variables, we thereby
arrive at
\bea
&&W_{p,q;m,n}(A,\theta)~=~\frac{1}{Z_{p,q}(A,\theta)}~\sum_{\vn,\vm}
\frac{(-1)^{|\vnu|}}{\prod_a\nu_a!}
{}~\int\limits_0^1\dd \mu ~ \e^{- 2 \pi \ii \mu\,p}
{}~\int\limits_0^1\dd \lambda ~ \e^{- 2 \pi \ii \lambda\,q}
\nn \\ &&~~~~~~\times\,\prod_{k=1}^{|\vnu| }
{}~\int\limits_{0^+}^\infty\frac{\dd z_k}{z_k }~\e^{-\frac{g^2A}2\,
z_k\bigl(m_k+\lambda+\theta\,(\mu-n_k)\bigr)^2}~
\e^{2\pi\ii z_k(\mu-n_k)}\nn  \\
&&~~~~~~ \times\,\sum_{l=1}^{|\vnu|}\frac{(z_l)^2}2~
\int\limits_{-1}^1\dd s~ \e^{-\pi\ii(1-s)\,n\,z_l}~
\e^{ -\frac{g^2 A}{4}\,z_l (1-s)\,\bigl[(m-n\,\theta)^2 + 2
(m-n\,\theta)\bigl(m_l + \lambda + \theta\,(\mu-n_l)\bigr) \bigr]} \non&& \ .
\eea

After a shift in $\lambda$ and a change of variable in the $s$-integral
we get
\bea
&&W_{p,q;m,n}(A,\theta)~=~\frac{1}{Z_{p,q}(A,\theta)}~
\sum_{\vn,\vm}\frac{(-1)^{|\vnu|}}{\prod_a\nu_a!}
{}~\int\limits_0^1\dd \mu ~ \e^{- 2 \pi \ii \mu\,(p-q\,\theta)}
{}~\int\limits_0^1\dd \lambda ~ \e^{- 2 \pi \ii \lambda\,q}
\nn  \\ &&~~~~~~\times\,\prod_{k=1}^{|\vnu| }
{}~\int\limits_{0^+}^\infty\frac{\dd z_k}{z_k }~\e^{-\frac{g^2A}2\,
z_k(m_k-n_k\,\theta+\lambda)^2}~\e^{2\pi\ii z_k(\mu-n_k)} \nn \\
&&~~~~~~ \times\,\sum_{l=1}^{|\vnu|}(z_l)^2\,
\int\limits_{0}^1\dd t~\e^{-2 \pi \ii t\,n\,z_l}~
\e^{ -\frac{g^2 A}{2}\,t\,z_l \bigl( (m-n\,\theta+ m_l-n_l\,
\theta+\lambda)^2 -(m_l-n_l\,\theta+\lambda)^2 \bigr)} \ ,
\label{resumncW}\eea
and by explicitly performing the integral over $t$ we arrive at the
final result
\bea
&&W_{p,q;m,n}(A,\theta)~=~\frac{1}{Z_{p,q}(A,\theta)}~
\sum_{\vn,\vm}\frac{(-1)^{|\vnu|}}{\prod_a\nu_a!}
{}~\int\limits_0^1\dd \mu ~ \e^{- 2 \pi \ii \mu\,(p-q\,\theta)}
{}~\int\limits_0^1\dd \lambda ~ \e^{- 2 \pi \ii \lambda\,q}
\nn  \\ &&~~~~~~~~~~\times\,\prod_{k=1}^{|\vnu| }
{}~\int\limits_{0^+}^\infty\frac{\dd z_k}{z_k }~\e^{-\frac{g^2A}2\,
z_k(m_k-n_k\,\theta+\lambda)^2}~\e^{2\pi\ii z_k(\mu-n_k)} \nn \\
&&~~~~~~~~~~\times\,\sum_{l=1}^{|\vnu|}z_l~
\frac{1-\e^{-2 \pi \ii  n\,z_l -\frac{g^2 A}{2}\,z_l\bigl( (m-n\,\theta)^2 + 2
(m-n\,\theta)(m_l-n_l\,\theta+ \lambda  )\bigr)}}
{2 \pi \ii  n+\frac{g^2 A}{2}\,\Bigl( (m-n\,\theta)^2 + 2
(m-n\,\theta)(m_l-n_l\,\theta+ \lambda  )\Bigr)} \ .
\label{corrstrongfinal}\eea
{}From (\ref{resumncW}) it is clear that the noncommutative electric
field variables $m-n\,\theta$ should be considered on the same footing as the
dual topological numbers $m_k-n_k\,\theta$. This is consistent with what
we have observed in section~\ref{OWLC} when we transformed the
correlation function in the commutative theory to one in the
noncommutative theory. Not only do the pairs of integers $(m,n)$ and
$(m_k,n_k)$ transform in the same manner under $SL(2,\zed)$ gauge
Morita equivalence, but they should also both be thought of as
algebraic numbers which are dual to topological numbers such as
$(p_k,q_k)$ that figure prominently in the definition of the correlation
function (\ref{wtheta}). The fact that the analog of the
cross-product
(\ref{crosspairing}) does not appear in (\ref{resumncW})
supports this identification.

In fact, the strong-coupling resummation (\ref{resumncW}) of the
pair correlation function gives a striking example of how
noncommutativity changes the definition of Yang-Mills theory on
the torus. As discussed in section~\ref{PhysCorrFns}, in the
commutative theory the winding number $m$ is related to group
theoretic quantities such as irreducible representations and the
highest weights $m_k$ which parameterize them. In the
noncommutative case, these row lengths are generalized to
two-dimensional objects $m_k-n_k\,\theta$ and in (\ref{resumncW})
there appears the analogous structure for the generalization of
the winding number $m$, the noncommutative electric field
$m-n\,\theta$.  The definition of a Wilson line in noncommutative
gauge theory thus does not depend solely on group theoretic
quantities but also on dipole momenta. The correlation function
(\ref{resumncW}) thereby gives an explicit example of how colour
and spacetime degrees of freedom are non-trivially tied together
in the noncommutative theory. The representation theory of the
noncommutative gauge group is determined by including the group of
translations on the torus with momentum integers into the usual
colour representation labels, which can no longer be trivially
separated. The twisting of the two group structures by
noncommutativity into a single Morita covariant structure is
strikingly similar to the realization of the noncommutative torus
as the crossed product algebra of functions on the circle ${\bf
S}^1$ by the integers $\zed$. The explicit mixing of the
translation group with the gauge symmetries of the model is a
manifestation of the teleparallelism property possessed by generic
noncommutative gauge theories~\cite{LangSz}. The arguments above
also justify the claims made in section~\ref{InstExp} that the
dual topological integers $(m_k,n_k)$ can be regarded as labelling
the contributions from virtual electric dipoles. This is an exact
realization of the proposal that the effective actions of generic
noncommutative quantum field theories can be expressed in terms of
open Wilson lines~\cite{rey3,rey4,Rey}.

There are various other salient features that can be readily deduced
from the strong-coupling form (\ref{resumncW}) of the open Wilson line
correlator. For example, the selection and weighting of partitions in
the instanton expansion (\ref{wtheta}) (or (\ref{wthetaint})) is
identical to that of the partition function
(\ref{Zpqpartitionsum}). Thus the arguments which established
$\theta$-smoothness of the vacuum amplitude~\cite{inprep} carry over
to the pair correlators. This is completely evident in the strong-coupling
resummation (\ref{resumncW}) (or (\ref{corrstrongfinal})), which shows
that the dipole amplitudes are analytic functions of the
noncommutativity parameter. Thus the poles in $\theta$ which sometimes arise
in
perturbative expansions~\cite{bnt,torr,GHLL}, and are
the hallmark of UV/IR mixing in noncommutative field theories, appear to be
non-perturbatively resummed into non-singular expressions in two-dimensional
noncommutative Yang-Mills theory. This supports suggestions that
the UV/IR mixing phenomenon is merely a resummable artifact of
perturbation theory.

\newsection{High-Energy Behaviour of Open Wilson Lines}

Because the extent of a dipole grows with its momentum, at high
energies the open Wilson lines become very long. Using this fact the
correlators of large momentum open Wilson lines can be analysed
perturbatively within a ladder approximation in the planar limit of
the gauge theory. For example, in the case of Yang-Mills theory on
noncommutative $\real^4$, the two-point functions of open Wilson line
operators at weak coupling exhibit a universal exponential growth with
high momentum as~\cite{ghi,rvr}
\beq
\Bigl\langle{\sf O}(k)\,{\sf O}(-k)\Bigr\rangle_{\rm 4D}\simeq\exp\left(
\sqrt{\frac{\sqrt{g^2N}~|k|\,|\theta\cdot k|}{4\pi}}~\right) \ .
\label{4Dhighmomlines}\eeq
This result implies an exponential suppression of higher correlation
functions at large momentum, except for those whereby a pair of dipole
momenta become anti-parallel in which case the exponential growth is
restored. The expression (\ref{4Dhighmomlines}) is identical to that
of the two-point correlation function computed using the supergravity
dual of noncommutative ${\cal N}=4$ supersymmetric Yang-Mills theory,
and it is reminiscent of the behaviour of high-energy fixed-angle
scattering amplitudes in string theory.

The open Wilson lines can be analysed on noncommutative $\real^2$ in a
particular decompactification limit of Yang-Mills theory on the
noncommutative torus \cite{bv}.
In the planar limit of the Morita dual theories
which approximate the given noncommutative gauge theory through a
sequence of rational approximants, i.e. $\theta=\lim_{r,s}r/s$, within
a certain regime one can find an analogous exponential increase in the
correlation functions with large momentum of the form
\beq
\Bigl\langle{\sf O}(k)\,{\sf O}(-k)\Bigr\rangle_{\rm 2D}\simeq
\frac{\exp\left(\,\frac{g^2A}2\,|k|\,N\,\right)}{\frac{g^2A}2\,
|k|\,N^2} \ , ~~ |k|<N-1 \ .
\label{2Dhighmomlines}\eeq
This expression is valid at strong 't~Hooft coupling, and is exact as
a function of the dipole momentum $|k|$, receiving corrections only at
order $\frac1N$. By redefining parameters it can be extrapolated to
the weak coupling result $\exp(\,\sqrt{g^2A\,N\,k^2\,\theta^2/\pi}~)$
obtained by resumming all planar ladder diagrams in perturbation
theory. Similar large momentum dependences have been observed
in~\cite{gsv,Bietenholz:2002ch}. In this final section we will use the exact
expressions for
the pair correlators obtained in this paper to study the high-energy
behaviour of the dipole scattering amplitudes.

Within the framework of gauge theory on the noncommutative torus, it
is in fact straightforward to prove that no such exponential rise with
momentum can occur. This fact is based on the integral representation
(\ref{wthetaint}) from which it can be shown that the correlator is
uniformly bounded as a function of the dimensionless Yang-Mills
coupling constant $g^2A$. For this, we use the elementary
integral inequality
\beq
\left|\,\frac{1}{2}\,\int\limits_{-1}^1\dd s~ \e^{ \pi  \ii a\,(1-s) -
b\,(1-s^2)}\,\right| \leq \frac{1}{2}\,\int\limits_{-1}^1\dd s~ \e^{ -
b\,(1-s^2)} \leq 1
\eeq
valid for any $a,b\in\real$, the definition (\ref{partitiondef}) of
partition, and the instanton expansion
(\ref{Zpqpartitionsum}) of the vacuum energy to obtain the bound
\bea
W_{p,q;m,n}(A,\theta)&\leq& \frac{1}{Z_{p,q}(A,\theta)}~
\sum_{\stackrel{\scriptstyle{\rm partitions}}
{\scriptstyle(\vp,\vq)}}\,\frac{(-1)^{|\vnu|}}{\prod_a\nu_a!}\,
\prod_{k=1}^{|\vnu|}\sqrt{\frac{2\pi^2}{g^2A\,
(p_k-q_k\,\theta)^3}}\non&&\times\,\exp\left[-\frac{2\pi^2}{g^2A}\,
\sum_{k=1}^{|\vnu|}\left(p_k-q_k\,\theta\right)
\left(\frac{q_k}{p_k-q_k\,\theta}-\frac q{p-q\,\theta}\right)^2\right]
\nn  \\&&\times\,\sum_{l=1}^{|\vnu|}
\left(p_l-q_l\,\theta\right)^2\non&&{~~}^{~~}_{~~}\non&\leq&
\frac{(p - q\,\theta)^2 }{Z_{p,q}(A,\theta)}~
\sum_{\stackrel{\scriptstyle{\rm partitions}}
{\scriptstyle(\vp,\vq)}}\,\frac{(-1)^{|\vnu|}}{\prod_a\nu_a!}\,
\prod_{k=1}^{|\vnu|}\sqrt{\frac{2\pi^2}{g^2A\,
(p_k-q_k\,\theta)^3}}\non&&\times\,\exp\left[-\frac{2\pi^2}{g^2A}\,
\sum_{k=1}^{|\vnu|}\left(p_k-q_k\,\theta\right)
\left(\frac{q_k}{p_k-q_k\,\theta}-\frac q{p-q\,\theta}\right)^2\right]
\non&&{~~}_{~~}^{~~}\non&=& (p-q\,\theta)^2 \ .
\label{Wbound}\eea
The lack of a runaway behaviour in momentum is a reflection of the
fact, established by the calculations of the present paper, that
noncommutative
Yang-Mills theory in two dimensions is a completely well-behaved
(finite) quantum field theory.

In fact, we can go even further and show that the correlation function
vanishes for high-energy dipoles. For this, we use the
strong-coupling expansion (\ref{corrstrongfinal}) in the limit of
large momentum $n$. For $\theta\neq0$ we find
\bea
\lim_{n\to\infty}\,W_{p,q;m,n}(A,\theta)&=& \frac{1}{Z_{p,q}(A,\theta)}~
\sum_{\vn,\vm}\frac{(-1)^{|\vnu|}}{\prod_a\nu_a!}
{}~\int\limits_0^1\dd\mu ~ \e^{- 2 \pi \ii \mu\,(p- q\,\theta)}
{}~\int\limits_0^1\dd\lambda ~ \e^{- 2 \pi \ii \lambda\,q}
\nn  \\ &&\times\,\prod_{k=1}^{|\vnu| }
{}~\int\limits_{0^+}^\infty\frac{\dd z_k}{z_k }~\e^{-\frac{g^2A}2\,
z_k(m_k-n_k\,\theta+\lambda)^2}~\e^{2\pi\ii z_k(\mu-n_k)}\non&&\times\,
\sum_{l=1}^{|\vnu|}~
\frac{2 z_l }{ g^2 A\,(m-n\,\theta)^2 } +O\Bigl((m-n\,\theta)^{-3}
\Bigr)\non&&{~~}_{~~}^{~~}
\nn \\& = & \frac{2\,(p-q\,\theta) }{ g^2 A\,(m-n\,\theta)^2 } +
O\Bigl((m-n\,\theta)^{-3}\Bigr) \ .
\eea
The vanishing of the pair correlator in this case is more rapid
than its commutative counterpart. For $\theta=0$,
the strong-coupling expansion
(\ref{corrstrongfinal}) can be reduced to the form
\bea
W_{p,q;m,n}(A,0)&=&\frac{1}{Z_{p,q}(A,0)}~\sum_{\stackrel{
\scriptstyle{\rm partitions}}{\scriptstyle\vp}}~\sum_{ \vm\in \zeds^p}
{}~\frac{(-1)^{|\vnu|}}{\prod_a\nu_a!}
{}~\int\limits_0^1 \dd \lambda ~ \e^{- 2 \pi \ii \lambda\,q}
\nn\\ &&\times\,\prod_{k=1}^{|\vnu|}\,
\frac{1}{p_k }~\e^{-\frac{g^2A}2\,p_k (m_k+\lambda )^2}~
\sum_{l=1}^{|\vnu|}\,p_l\non&&\times\,\sum_{a=1}^p~
\frac{1-\e^{  -\frac{g^2 A}{2}\,p_l\bigl(m^2 + 2m\,(m_a+ \lambda )\bigr)}}
{2 \pi \ii  n+\frac{g^2 A}{2}\,\Bigl(m^2 + 2m\,(m_a+ \lambda )\Bigr)} \ ,
\eea
which vanishes as $\frac1n$ for $n\to\infty$.

Even in the decompactification limit whereby the theory is projected out
into gauge theory on the noncommutative plane, such momentum behaviour
does not occur. This is because, like in the case of the partition
function, the large area limit of (\ref{corrstrongfinal}) is trivial,
\beq
\lim_{g^2A\to\infty}\,W_{p,q;m,n}(A,\theta)=0 \ .
\label{largearea}\eeq
This result is most easily derived from the integral representation
(\ref{wthetaint}). Note that this vanishing of the momentum representation
correlator at $\theta=0$ is consistent with the limit of the position
representation in (\ref{Wp1largeA2}) in the physical theory, due to the fact
that the Fourier transforms (\ref{Fourierloop}) and (\ref{physcorrrecover}) do
not commute with taking the large area limit. Thus the momentum representation
correlator can be trivial at large $A$ without contradicting the non-triviality
of the position representation correlator.

Consequently, the correlation functions of infinitely-long,
anti-parallel open Wilson lines on the noncommutative plane
vanish, and this is a reflection of the fact that the star-gauge
invariance in this model contains area-preserving
diffeomorphisms~\cite{inprep}. Thus the long (high momentum) dipole excitations
in
the present case play no role and are completely decoupled from the
dynamics of the noncommutative gauge theory. This conclusion holds in
the strong coupling regime of the gauge theory whereby the connections with
string theory are expected to hold. However, this does not mean that
the noncommutative field theory is free from stringy effects.
For example, we have explicitly demonstrated in (\ref{Zpqresumfinal}) and
(\ref{corrstrongfinal}) that the observables of the system can
expressed in terms of the non-local dipole quanta, with non-vanishing
correlations at intermediate energy scales. Furthermore, we must also
remember that the results (\ref{2Dhighmomlines}) are derived in the planar
$N\to\infty$ limit. The appropriate large $N$ limit of gauge theory on the
noncommutative torus may well agree with these predictions.

\subsection*{Acknowledgements}

We thank J.~Ambj\o rn, W.~Bietenholz, J.~Nishimura, N.~Obers and
S.~Waldmann for helpful discussions. The work of R.J.S. was supported
in part by an Advanced Fellowship from the Particle Physics and
Astronomy Research Council~(U.K.).

\end{document}